\documentclass{aastex631}
\usepackage{amsmath}
\usepackage{soul}
\usepackage{graphicx}
\usepackage{subfigure}
\usepackage{fourier}
\usepackage{arcs}
\usepackage{url}
\usepackage{color}
\usepackage{amsbsy}
\usepackage{mathrsfs}
\usepackage{bm}
\usepackage{booktabs}
\usepackage{multirow}
\usepackage{url}
\usepackage{ulem}
\usepackage{makecell}
\usepackage{media9}
\usepackage{natbib}
\usepackage{tikz}
\begin{document}

\title{COCONUT: A coronal model with an energy decomposition strategy}

\author[0000-0002-4217-6990]{Hao P. Wang}
\affiliation{Centre for Mathematical Plasma-Astrophysics, Department of Mathematics, KU Leuven, Celestijnenlaan 200B, 3001 Leuven, Belgium; \url{Stefaan.Poedts@kuleuven.be}; \url{andrea.lani@kuleuven.be}; \url{haopeng.wang1@kuleuven.be}}

\author[0000-0002-1743-0651]{S. Poedts}
\affiliation{Centre for Mathematical Plasma-Astrophysics, Department of Mathematics, KU Leuven, Celestijnenlaan 200B, 3001 Leuven, Belgium}
\affiliation{Institute of Physics, University of Maria Curie-Skłodowska, ul. Radziszewskiego 10, 20-031 Lublin, Poland}

\author[0000-0003-4017-215X]{A. Lani}
\affiliation{Centre for Mathematical Plasma-Astrophysics, Department of Mathematics, KU Leuven, Celestijnenlaan 200B, 3001 Leuven, Belgium}
\affiliation{Von Karman Institute For Fluid Dynamics, Waterloosesteenweg 72, 1640 Sint-Genesius-Rode, Brussels, Belgium}

\author[0000-0003-3670-4678]{R. Dhib}
\affiliation{Centre for Mathematical Plasma-Astrophysics, Department of Mathematics, KU Leuven, Celestijnenlaan 200B, 3001 Leuven, Belgium}

\author[0000-0002-4014-1815]{L. Linan}
\affiliation{Centre for Mathematical Plasma-Astrophysics, Department of Mathematics, KU Leuven, Celestijnenlaan 200B, 3001 Leuven, Belgium}

\author[0000-0002-1986-4496]{T. Baratashvili}
\affiliation{Centre for Mathematical Plasma-Astrophysics, Department of Mathematics, KU Leuven, Celestijnenlaan 200B, 3001 Leuven, Belgium}

\author[0000-0003-4616-947X]{H.-J. Jeong}
\affiliation{Centre for Mathematical Plasma-Astrophysics, Department of Mathematics, KU Leuven, Celestijnenlaan 200B, 3001 Leuven, Belgium}
\affiliation{School of Space Research, Kyung Hee University, Yongin, 17104, Republic of Korea}

\author[0000-0002-4391-393X]{Yu H. Zhou}
\affiliation{School of Astronomy and Space Science and Key Laboratory of Modern Astronomy and Astrophysics, Nanjing University, Nanjing 210023, China}

\author{Yu C. Li}
\affiliation{Centre for Mathematical Plasma-Astrophysics, Department of Mathematics, KU Leuven, Celestijnenlaan 200B, 3001 Leuven, Belgium}

\author[0009-0008-0922-3995]{M. Najafi-Ziyazi }
\affiliation{Centre for Mathematical Plasma-Astrophysics, Department of Mathematics, KU Leuven, Celestijnenlaan 200B, 3001 Leuven, Belgium}

\author{J. Wang}
\affiliation{School of Systems Science, Beijing Normal University, Beijing 100875, China}

\author[0000-0003-3364-9183]{B. Schmieder}
\affiliation{Centre for Mathematical Plasma-Astrophysics, Department of Mathematics, KU Leuven, Celestijnenlaan 200B, 3001 Leuven, Belgium}
\affiliation{Observatoire de Paris, LIRA, UMR8254 (CNRS), F-92195 Meudon Principal Cedex, France}
\affiliation{LUNEX EMMESI COSPAR-PEX Eurospacehub, Kapteyn straat 1, Noordwijk 2201 BB, Netherlands}

\author[0000-0002-9865-5245]{W. S. Wang}
\affiliation{CAS Key Laboratory of Geospace Environment, Department of Geophysics and Planetary Sciences, University of Science and Technology of China, Hefei, 230026, People's Republic of China}

\author[0000-0002-1349-3663]{E. Husidic}
\affiliation{Centre for Mathematical Plasma-Astrophysics, Department of Mathematics, KU Leuven, Celestijnenlaan 200B, 3001 Leuven, Belgium}
\affiliation{Department of Physics and Astronomy, University of Turku, 20014 Turku, Finland}

\begin{abstract}

In this paper, we propose an energy decomposition method combined with an HLL Riemann solver that includes an additional dissipation term in the energy equation to improve the numerical stability of the fully implicit, time-evolving coronal model COCONUT and extend its applicability to solar-maximum phases. In MHD simulations that evolve conservative variables in time, the thermal pressure is typically computed by subtracting the magnetic and kinetic energies from the total energy. In low-$\beta$ (the ratio of thermal to magnetic pressure; $< 10^{-3}$) regions, discretization errors of magnetic energy can be comparable to the thermal pressure, potentially leading to negative thermal pressure and causing the simulation to crash. Therefore, we update the decomposed energy, excluding the magnetic energy, at each time step. It avoids subtracting a large magnetic energy from the total energy to obtain a very small thermal pressure in low-$\beta$ regions, thereby improving the numerical stability of MHD models. We validate the algorithm using a time-evolving solar-maximum Carrington rotation simulation in 2025, which the previous code failed to run to completion. We also perform quasi-steady-state coronal simulations and 2D benchmark tests to further assess the algorithm's performance. The simulation results show that the algorithm produces results nearly identical to those obtained using the traditional full energy equation during solar minimum, while significantly improving COCONUT's ability to simulate coronal evolution under strong magnetic fields, even including fields exceeding 100 Gauss with $\beta<10^{-3}$. This method provides a promising approach for performing quasi-realistic coronal simulations during solar maxima.

\end{abstract}

\section{Introduction}\label{sec:intro}
Space weather can impact the performance and reliability of both space- and ground-based technological systems, alongside posing risks to human life and health. There is an urgent need to develop advanced Sun-to-Earth model chains \cite[e.g.][]{Feng_2011Chinese,Feng_2013Chinese,Pomoell2018020,Poedts_2020,Hayashi_2021} to better understand the mechanisms of space weather and ultimately provide reliable space weather forecasts hours to days in advance. One of the promising methods to achieve this goal is developing efficient and reliable magnetohydrodynamic (MHD) solar-terrestrial models \cite[e.g.][and references therein]{Owens2017,Feng2020book}. 
Given that coronal models are typically the most complex and computationally intensive component, and are essential for initializing other modules and accurately simulating solar disturbances such as coronal mass ejections \cite[CMEs;][]{Yang_2021,wang2025sipifvmobservationbasedmagnetohydrodynamicmodel,WangSIPtheoriticalCME}, one of the primary challenges in improving the overall efficiency and reliability of the Sun-to-Earth model chains is developing MHD coronal models that simultaneously maintain high computational efficiency, strong numerical stability, and required accuracy \citep{Feng_2021,Brchnelova_2022,Wang_2022,Wang2022_CJG,Wang2025_FirsttimeevolvingCOCONUT,wang2025COCONUTMayEvent,wang2025sipifvmtimeevolvingcoronalmodel}.

To develop advanced MHD coronal models suitable for practical applications, such as daily space weather forecasting, it is essential to first address the low-$\beta$ ($\beta$ is the thermal to magnetic pressure ratio) issues. Near the solar surface, the magnetic field around active regions can be very strong, and the plasma $\beta$ can drop as low as $10^{-4}$ \citep[e.g.][]{Bourdin2017}, which brings about the first successful study of low plasma $\beta$ coronal environment modeling \citep{Feng_2021,Wang_2022}. Around such regions, extremely fast Alfv{\'e}nic or fast
magnetosonic waves usually result in extremely small time steps due to the Courant–Friedrichs–Lewy (CFL) stability condition, which limits the time step to not exceed the spatial mesh size divided by the fastest wave speed, leading to very low computational efficiency. 
To improve the computational efficiency of coronal simulations involving low-$\beta$ regions, implicit temporal integration methods are a promising approach. In recent years, implicit methods that enable selecting large time steps exceeding the CFL stability limitation have been demonstrated to be able to perform several tens to over a thousand times faster than explicit models in quasi-steady-state background coronal simulations \citep{WANG2019181,Wang_2022,Wang2022_CJG,Perri_2022,Perri_2023}. By selecting approximate large time steps combined with additional iterations within each physical time step, implicit coronal models can be both time-accurate and much more computationally efficient than explicit models in both time-evolving coronal simulations \citep{Wang2025_FirsttimeevolvingCOCONUT,wang2025COCONUTMayEvent,wang2025sipifvmtimeevolvingcoronalmodel} and CME simulations \citep{Linan_2023,Linan_2025,wang2025sipifvmobservationbasedmagnetohydrodynamicmodel,WangSIPtheoriticalCME}.

Another challenge associated with low-$\beta$ conditions is that MHD models often suffer from poor numerical stability in such regions. In MHD simulations that advance discretized conservative variables in time, thermal pressure $p$ is typically obtained by subtracting the magnetic energy $\left|\mathbf{B}\right|^2\big/2$ and the kinetic energy $\rho \left|\mathbf{v}\right|^2\big/2$ from the total energy $E=p\big/\left(\gamma-1\right)+\left|\mathbf{B}\right|^2\big/2+\rho \left|\mathbf{v}\right|^2\big/2$, calculated as $p=\left(\gamma-1\right)\left(E-\left|\mathbf{B}\right|^2\big/2-\rho \left|\mathbf{v}\right|^2\big/2\right)$. In low-$\beta$ regions, discretization errors in the magnetic energy can be comparable to the magnitude of the thermal pressure. As a result, calculating thermal pressure by subtracting a much larger magnetic energy from the total energy can yield non-physical negative values, causing the code to crash.

In recent decades, although significant progress has been made in addressing the low-$\beta$ issue described above, most approaches remain limited to resolving one aspect while compromising the performance of another. 
For example, adopting simplified 1~D equations for plasma motion below $1.1\;R_s$ significantly improves the computational efficiency of the original 3~D Alfv{\'e}n Wave Solar atmosphere Model (AWSoM) coronal model \citep{Jin_2017, Sokolov2021}. Similarly, using an empirical magnetofrictional model below 1.15$\;R_s$ improves the numerical stability and efficiency of the global coronal–heliospheric MHD (GHM) model \citep{Hoeksema2020, Hayashi_2021}. Although these simplifications help avoid low-$\beta$ issues, they also lead to a significant loss of fidelity in the simulated coronal structures in the low solar atmosphere. 
Additionally, the AWSoM and Magnetohydrodynamic Algorithm outside a Sphere (MAS) coronal models include an artificially broadened transition region with the plasma density significantly increazed, and employ extremely high-resolution meshes, with cell sizes below 1~Mm in the radial direction near the solar surface \citep{Mok_2005,Lionello_2008,Mikic_2013,MIKIC2018NatA,van_der_Holst_2022}. This novel approach also improves the numerical stability near the solar surface.
A more detailed description of these models is available in \cite{Wang2025_FirsttimeevolvingCOCONUT,wang2025COCONUTMayEvent,wang2025sipifvmtimeevolvingcoronalmodel}.

Alternatively, some researchers split the magnetic field $\mathbf{B}$ into a time-invariant potential \cite[e.g.][]{Tanaka1995,Powell1999,FUCHS2010JCP,Guo2015} or an arbitrary background magnetic field $\mathbf{B}_0$ \citep{Xia_2018}, and into a time-dependent field $\mathbf{B}_1$, and update $\mathbf{B}_1$ instead of $\mathbf{B}$ during coronal simulations \cite[e.g.][]{Feng_2010,Yang2012,Licaixia2018,WANG201967,Feng_2021,Liu_2023,Wang_2022,Wang2022_CJG,WangSIPtheoriticalCME}. This magnetic field decomposition method performs well for quasi-steady-state coronal simulations constrained by a single static magnetogram, with $\mathbf{B}_1$ remaining small during the simulations. \cite{wang2025sipifvmtimeevolvingcoronalmodel} further proposed the extended magnetic field decomposition method to maintain $\mathbf{B}_1$ small during the time-evolving simulations by introducing an additional temporally piecewise constant field to accommodate part of the non-potential magnetic field, and implemented it in the solar interplanetary phenomena implicit finite volume method (SIP-IFVM) coronal model. The effectiveness of this approach has also been validated in an observation-based CME simulation \citep{wang2025sipifvmobservationbasedmagnetohydrodynamicmodel}. However, this extended magnetic field decomposition method introduces an additional variable, making its integration into an existing implicit solver for coronal plasmas with the complexity of coronal model COCONUT \citep[COolfluid COroNal UnsTructured;][]{Perri_2022,Perri_2023,Kuzma_2023,Baratashvili,Wang2025_FirsttimeevolvingCOCONUT,wang2025COCONUTMayEvent} non-trivial.

In addition, some researchers directly update the thermal pressure in CME simulations within the coronal and inner heliospheric regions \citep{Shen2007,Shen2009,Shen2011,Zhou_2018,Liu_2019,Guo_2025}. 
Although updating the thermal pressure independently compromises the conservative property of the governing equations and may lead to discrepancies with the total energy evolution, which includes the kinetic and magnetic energy governed by the momentum and induction equations, this approach helps prevent the occurrence of non-physical negative thermal pressures in low-$\beta$ regions resulting from catastrophic cancellation. While preserving the conservative form of the governing equations is essential to maintain the model consistency \citep{van_der_Holst_2014}, it is sometimes necessary to compromise this requirement to achieve the required numerical stability \citep{BERNADES2023112477}. 

To balance numerical stability with the conservative property of the energy equation, some researchers update the decomposed energy, which includes internal and kinetic energy, in MHD simulations of the inner heliosphere, as well as Earth's and planetary magnetospheres \citep{LYON20041333,Merkin2016,Zhang_2019,Provornikova2024}. 
Although non-conservative forms of the energy equation can sometimes lead to nonphysical results when non-ideal processes such as shocks occur, they still satisfy the Rankine–Hugoniot relations within numerical truncation error \citep{LYON20041333, Zhang_2019}. It has been reported that, although the truncation-level error is not as small as the round-off error associated with the total energy formulation, it remains sufficient to ensure that shock solutions are accurate to a good approximation \citep{LYON20041333}.  In addition to the decomposed energy equation strategy, they adopted constrained transport method \citep{EvansandHawley1988,DEVORE1991142} to maintain the divergence-free magnetic field, utilized the semi-relativistic (Boris) correction with an artificially reduced speed of light \citep{Boris1970,GOMBOSI2002176} to alleviate the extremely small time-step limitation in the low-$\beta$ region, and calculated the gas-hydro flux and magneto-hydro flux separately \citep{Zhang_2019}. In this paper, we further derive the quasi-conservative decomposed energy equation by subtracting the magnetic energy equation from the total energy equation and for the first time to implement it in MHD coronal simulations. Additionally, to mitigate the drawbacks of the decomposed energy equation in handling non-ideal processes, which may produce nonphysical results and cause the code to crash, we introduce an approximate additional dissipation term into the energy component of the Harten-Lax-van Leer (HLL) Riemann solver \citep{Feng_2021,Wang2025_FirsttimeevolvingCOCONUT} used in our code.  

This paper aims to balance computational efficiency, numerical stability, and temporal accuracy in coronal simulations by implementing the decomposed energy approach, combined with an HLL Riemann solver that incorporates an additional dissipation term in the energy equation, in the time-evolving coronal model COCONUT \citep{wang2025COCONUTMayEvent}. This method will further enable COCONUT to perform comparable to the implicit SIP-IFVM MHD coronal model, which employs an extended magnetic field decomposition method to maintain numerical stability in solving time-varying low-$\beta$ issues \citep{wang2025sipifvmobservationbasedmagnetohydrodynamicmodel,wang2025sipifvmtimeevolvingcoronalmodel}. 
 
Based on the above considerations, the paper is organized as follows. In Section \ref{NumericalAlgorithm}, we introduce the numerical algorithm used to make the coronal simulations and describe the derivation of the decomposed energy equation used to improve the numerical stability of the MHD coronal model. In Section~\ref{sec:Numerical Results}, we present the time-evolving simulation results around the solar maximum CR and compare them with coronal observations. 
This section also presents a comparison of quasi–steady-state simulation results obtained with both versions of the COCONUT model, using the total and decomposed energy equations, during a solar minimum and an increasing activity phase, to further validate the energy decomposition method. In Appendix~\ref{sec:benchmark}, we present the results of the Orszag–Tang vortex problem and the MHD rotor problem \citep{Stone_2008,CIUCA2020100042} to further validate and analyze the decomposed energy equation method and the HLL Riemann solver with an additional dissipation term in the energy component.
In Section~\ref{sec:Conclusion}, we summarize the features of the implicit, time-evolving coronal model COCONUT that adopts the decomposed energy equation and provide concluding remarks. 

\section{Numerical algorithm}\label{NumericalAlgorithm}
COCONUT is a recently developed implicit MHD coronal model built on the Computational Object-Oriented Libraries for Fluid Dynamics (COOLFluiD) framework \citep{lani1,lani13,kimpe2}\footnote{\url{https://github.com/andrealani/COOLFluiD/wiki}}. The adoption of implicit temporal integration enables high computational efficiency. The use of an unstructured grid mesh provides flexibility in mesh division and allows the model to avoid degeneracy near the poles. The decomposed energy strategy proposed in this paper further ensures COCONUT to remain numerically stable during solar maximum coronal simulations. 
It employs the Generalized Minimal Residual (GMRES) method \citep{Saad1986}, an iterative algorithm that approximates the solution of a given linear system by minimizing the residual over a Krylov subspace, combined with a restricted additive Schwarz preconditioner to solve the linearized system \citep{Lani2008AnOO}, as provided in the Portable, Extensible Toolkit for Scientific computation \citep[PETSc;][]{petscwebpage}. By adopting a large time step exceeding the CFL stability limitation, the quasi-steady-state version COCONUT using a backward Euler method achieved a speed up of 35X \citep{Perri_2022,Perri_2023} compared with WindPredict \citep{Perri2018SimulationsOS,Parenti_2022}, an explicit coronal model based on the PLUTO code \citep{Mignone_2007}. 
\citet{wang2025COCONUTMayEvent,Wang2025_FirsttimeevolvingCOCONUT} further extended the model to a time-evolving version and adopted cubic Hermite interpolation, using four adjacent observation-based synoptic magnetograms as a stencil, to construct time-evolving magnetograms at each physical time step. To maintain the required temporal accuracy, a second-order backward differentiation formula (BDF2), with Newton iterations performed within each time step, is employed for temporal integration. 
As in \citet{Wang2025_FirsttimeevolvingCOCONUT} and \citet{wang2025COCONUTMayEvent}, the Newton iterations at each physical time step are terminated and the solution is advanced to the next physical time step when the $L_2$ norm of the difference in the solution vectors between two successive iterations falls below $10^{-3}$, or when a maximum of ten sub-iterations is reached.

In this paper, as usual, for the time-evolving simulations, the same BDF2 with Newton iterations applied at each physical time step is used for temporal integration, whereas the backward Euler method is adopted for the quasi-steady-state coronal simulations. Meanwhile, the boundary conditions described in \citet{Wang2025_FirsttimeevolvingCOCONUT,wang2025COCONUTMayEvent} are employed. Additionally, the numerical convective (inviscid) flux across cell interfaces is calculated using the HLL Riemann solver \citep{Harten1983,Feng2020book,Wang_2022}, with an appropriate dissipation term incorporated into the energy equation component in low-$\beta$ regions to improve numerical stability.
Although the energy equation is formulated in decomposed energy form, this approach does not alter the original MHD governing equations. Therefore, we retain the same eigenvalues and eigenvectors as in the previous versions of COCONUT that solve the full energy equation \citep{Wang2025_FirsttimeevolvingCOCONUT}. The positivity-preserving (PP) measures, as implemented for the thermal pressure and plasma density described in \citet{wang2025COCONUTMayEvent}, are also adopted in this paper. The primary distinction from previous work is the implementation of the decomposed energy equation.

\subsection{Governing equations and grid system}\label{Governingequations} 
In this paper, we drive the time-evolving thermodynamic MHD coronal model by a series of time-varying magnetograms to simulate the coronal evolution in an inertial coordinate system. The governing equation is defined similarly to that in \cite{Wang2025_FirsttimeevolvingCOCONUT} and \cite{wang2025COCONUTMayEvent}, but adopts a decomposed energy approach. It can be described in the following form:
\begin{equation}\label{MHDinsolarwind}
\frac{\partial \mathbf{U}}{\partial t}+\nabla \cdot \mathbf{F}\left(\mathbf{U}\right)=\mathbf{S}\left(\mathbf{U},\nabla \mathbf{U}\right) \,,
\end{equation}
where $t$ is the time variable, $\mathbf{U}=\left(\rho, \rho \mathbf{v}, \mathbf{B}, E_1, \psi\right)^T$ represents the vector of conservative variables, $\nabla \mathbf{U}$ denotes the spatial derivative of $\mathbf{U}$, and 
\begin{equation}\label{InviscidFluxDecomposedenergy}
    \mathbf{F}\left(\mathbf{U}\right)=
                     \begin{pmatrix}
\rho \mathbf{v}  \\
\rho\mathbf{v}\mathbf{v}+p_{T}\mathbf{I}-\mathbf{B}\mathbf{B}\\
\mathbf{vB}-\mathbf{Bv}+\psi\mathbf{I}\\
\left(E_1+p\right)\mathbf{v}\\
V_{\rm ref}^2 \mathbf{B}
                     \end{pmatrix}\,,
\end{equation}
is the inviscid flux vector.
Here, $\mathbf{B}=\left(B_x,B_y,B_z\right)$ and $\mathbf{v}=\left(u,v,w\right)$ denote the magnetic field and velocity in the Cartesian coordinate system, $E_1=p\big/\left(\gamma-1\right)+\rho \mathbf{v}^2\big/2$ means the decomposed energy density, and $\gamma=5\big/3$ is the adiabatic index. $\mathbf{B}\mathbf{v}\equiv \mathbf{B} \otimes \mathbf{v}$, $\mathbf{v}\mathbf{B}\equiv \mathbf{v} \otimes \mathbf{B}$, $\mathbf{B}\mathbf{B}\equiv \mathbf{B} \otimes \mathbf{B}$, and $\mathbf{v}\mathbf{v}\equiv \mathbf{v} \otimes \mathbf{v}$ with $\otimes$ denoting the dyadic (tensor) products of two corresponding vectors. For example,
$$
\mathbf{B} \otimes \mathbf{v} =
\begin{bmatrix}
B_x u & B_x v & B_x w \\
B_y u & B_y v & B_y w \\
B_z u & B_z v & B_z w
\end{bmatrix}\,.
$$ Additionally, $\rho$ and $p$ represent the plasma density and thermal pressure, respectively. For convenience, a factor of $1\big/\sqrt{\mu_0}$, where $\mu_0 = 4 \times 10^{-7} \pi ~ \rm H \cdot m^{-1}$ is the magnetic permeability, is absorbed into the definition of the magnetic field. Moreover, $\psi$ and $V_{\rm ref}$ denote the Lagrange multiplier and the reference propagation speed of the numerical divergence error defined in the hyperbolic generalized Lagrange multiplier (HGLM) method \citep{Dedner2002JCoPh,YALIM20116136}. 
In this paper, we set 
$V_{\rm ref} = 3\, V_{a,\rm ref}$, where
$V_{a,\rm ref} = \frac{B_{\rm ref}}{\sqrt{\rho_{\rm ref}}}$ with $\rho_{\rm ref} = 1.67 \times 10^{-13}~\rm kg~m^{-3}$ and 
$B_{\rm ref} = 2.2 \times 10^{-4}~\rm T$ denoting the characteristic plasma density and magnetic field strength at the solar surface that are used to normalize physical variables to code units \citep{Wang2025_FirsttimeevolvingCOCONUT}. Meanwhile, $\mathbf{S}\left(\mathbf{U},\nabla \mathbf{U}\right)=\mathbf{S}_{\rm gra}+\mathbf{S}_{\rm heat}+\mathbf{S}_{\rm{DECOMP}}$ represents the vector of the source term corresponding to the gravitational force, the heating source terms, and the source term resulting from the decomposed energy equation.  
The radiative loss term, the coronal heating term, and the thermal conduction term are included in the heating source term $\mathbf{S}_{\rm heat}$. In this paper, $\mathbf{S}_{\rm heat}$ and $\mathbf{S}_{\rm gra}$ are defined the same as in \cite{wang2025COCONUTMayEvent}. The vector source term caused by the decomposed energy method can be written as
\begin{equation}\label{SourceTermDecmpEnergy}
\mathbf{S}_{\rm{DECOMP}}=\left(0, \mathbf{0}, \mathbf{0},-~\mathbf{B}\cdot\left(\mathbf{v}\cdot\nabla\mathbf{B}\right)+\mathbf{v}\cdot\left(\mathbf{B}\cdot\nabla\mathbf{B}\right),0\right)^{T}.
\end{equation}

The computational domain is a spherical shell extending from 1.01 to approximately 25~$R_s$. It is discretized into the unstructured 6$_{th}$-level subdivided geodesic mesh \citep{Brchnelova2022}, resulting in 1,515,520 non-overlapping prismatic cells. In the radial direction, the grid comprises 74 progressively stretched layers, each containing 20,480 cells, with an angular resolution of approximately $1.8^\circ$. Considering that this angle corresponds to about 200 minutes for the Sun to rotate through, and the implicit temporal integration adopted in COCONUT allows flexibility in selecting a large time step, we adopt a fixed time step of 5 minutes, which is sufficient to resolve the evolution of the primary coronal structures and maintain desired computational efficiency \citep{Wang2025_FirsttimeevolvingCOCONUT,wang2025COCONUTMayEvent}.

\subsection{Derivation of the energy decomposition method}\label{DerivationofthedecomposedMHDequations}
Considering that $\left(\mathbf{B}+\epsilon~\mathbf{B}\right)^2-\mathbf{B}^2\equiv 2~\epsilon~\mathbf{B}^2+\epsilon^2~\mathbf{B}^2$ with $\epsilon~\mathbf{B}$ denoting the discretization error of the magnetic field $\mathbf{B}$, the discretization error of the magnetic pressure can be comparable to the thermal pressure in low $\beta$ regions. This situation can result in a nonphysical negative thermal pressure when the thermal pressure is derived from the energy density in MHD coronal simulations.
To avoid such undesirable situations, we propose an energy decomposition method in which the total energy $E$ is split into the magnetic energy $\mathbf{B}^2\big/2$ and the remaining part $E_1=p\big/\left(\gamma-1\right)+\rho \mathbf{v}^2\big/2$. In the decomposed energy conservation equation, we update $E_1$ instead of $E$ at each physical time step, without altering the analytical form of the MHD equations. This strategy is analogous to a special case of the extended magnetic field decomposition approach proposed by \cite{wang2025sipifvmtimeevolvingcoronalmodel}, in which the time-evolving magnetic field component $\mathbf{B}_1$ is reset to zero after the solutions are updated at each physical time step \citep{wang2025sipifvmobservationbasedmagnetohydrodynamicmodel}. In what follows, we describe the derivation of the energy decomposition equation.

The original energy equation is described as
\begin{equation}\label{orienergyequation}
\frac{\partial E}{\partial t}+\nabla \cdot\left[\left(E+p+\frac{\mathbf{B}^2}{2}\right)\mathbf{v}-\mathbf{B}\left(\mathbf{v}\cdot\mathbf{B}\right)\right]
=-\left(\mathbf{v} \cdot \mathbf{B}\right)\left(\nabla \cdot \mathbf{B}\right)\,,
\end{equation}
with the total energy $E$ described as 
\begin{equation}\label{TotalEnergyE}
E=\frac{1}{2}\mathbf{B}^{2}+E_1\,.
\end{equation}
From Eq.~(\ref{TotalEnergyE}), we get 
\begin{equation}\label{EnergyEquation}
\frac{\partial E}{\partial t}=\frac{\partial E_1}{\partial t}+ \mathbf{B}\cdot \frac{\partial \mathbf{B}}{\partial t}\,.
\end{equation}
The induction equation is described as
\begin{equation}\label{oriinductionequation}
\frac{\partial \mathbf{B}}{\partial t}+\nabla \cdot\left(\mathbf{v \, B}-\mathbf{B \, v}\right)=-\left(\nabla \cdot \mathbf{B}\right)\mathbf{v}\,.
\end{equation}
Multiplying both the left- and right-hand sides of Eq.~(\ref{oriinductionequation}) by $\mathbf{B}$, we have
\begin{equation}\label{B0dotPartialB1_ori}
\mathbf{B} \cdot \frac{\partial \mathbf{B}}{\partial t} = -\nabla \cdot \left( \mathbf{v} \, \mathbf{B} - \mathbf{B} \, \mathbf{v} \right) \cdot \mathbf{B} - \left(\mathbf{v} \cdot \mathbf{B}\right)\left(\nabla \cdot \mathbf{B}\right)\,.
\end{equation}
Considering that
$$
\begin{aligned}
\left[\nabla \cdot \left( \mathbf{v} \, \mathbf{B}\right)\right]\cdot \mathbf{B}=&\left( \nabla \cdot \mathbf{v}\right) \left(\mathbf{B} \cdot \mathbf{B}\right)+ \left(\mathbf{v} \cdot \nabla \mathbf{B}\right) \cdot \mathbf{B} \\= &\nabla \cdot \left[ \mathbf{v} \, \left(\mathbf{B} \cdot \mathbf{B}\right)\right]
-\mathbf{v} \cdot \nabla \left(\mathbf{B}  \cdot \mathbf{B}\right)+\left(\mathbf{v} \cdot \nabla \mathbf{B}\right) \cdot \mathbf{B}\,,
\end{aligned}$$
and
$$
\begin{aligned}
\left[\nabla \cdot \left( \mathbf{B} \, \mathbf{v}\right)\right]\cdot \mathbf{B}=&\left( \nabla \cdot \mathbf{B}\right) \left(\mathbf{v} \cdot \mathbf{B}\right)+ \left(\mathbf{B} \cdot \nabla \mathbf{v}\right) \cdot \mathbf{B} \\
=& \nabla \cdot \left[ \mathbf{B} \, \left(\mathbf{v} \cdot \mathbf{B}\right)\right]
-\mathbf{B} \cdot \nabla \left(\mathbf{B} \cdot \mathbf{v}\right)+\left(\mathbf{B} \cdot \nabla \mathbf{v}\right) \cdot \mathbf{B}\,,
\end{aligned}$$
Eq.~(\ref{B0dotPartialB1_ori}) is equivalent to
\begin{equation}\label{B0dotPartialB1}
\mathbf{B} \cdot \frac{\partial \mathbf{B}}{\partial t} = -\nabla \cdot \left[ \left(\mathbf{B} \cdot \mathbf{B}\right) \, \mathbf{v}  - \left(\mathbf{v} \cdot \mathbf{B}\right) \, \mathbf{B} \right] - \left(\mathbf{v} \cdot \mathbf{B}\right)\left(\nabla \cdot \mathbf{B}\right)+\mathbf{v} \cdot \nabla \left(\mathbf{B}  \cdot \mathbf{B}\right)-\left(\mathbf{v} \cdot \nabla \mathbf{B}\right) \cdot \mathbf{B}-\mathbf{B} \cdot \nabla \left(\mathbf{B} \cdot \mathbf{v}\right)+\left(\mathbf{B} \cdot \nabla \mathbf{v}\right) \cdot \mathbf{B}\,.
\end{equation}
From Eq.~(\ref{B0dotPartialB1}) and Eq.~(\ref{EnergyEquation}) we obtain
\begin{equation}\label{partialE}
\frac{\partial E}{\partial t} = \frac{\partial E_1}{\partial t} - \nabla \cdot \left[ \left(\mathbf{B} \cdot \mathbf{B}\right) \, \mathbf{v}  - \left(\mathbf{v} \cdot \mathbf{B} \right) \, \mathbf{B} \right] - \left(\mathbf{v} \cdot \mathbf{B}\right)\left(\nabla \cdot \mathbf{B}\right)+\mathbf{v} \cdot \nabla \left(\mathbf{B}  \cdot \mathbf{B}\right)-\left(\mathbf{v} \cdot \nabla \mathbf{B}\right) \cdot \mathbf{B}-\mathbf{B} \cdot \nabla \left(\mathbf{B} \cdot \mathbf{v}\right)+\left(\mathbf{B} \cdot \nabla \mathbf{v}\right) \cdot \mathbf{B}\,.
\end{equation}
From Eq.~(\ref{orienergyequation}) and Eq.~(\ref{TotalEnergyE}) we obtain
\begin{equation}\label{partialEandB1}
\frac{\partial E}{\partial t} = -\nabla \cdot \left[ \left( E_1 + p + \mathbf{B}^2 \right) \mathbf{v} - (\mathbf{v} \cdot \mathbf{B}) \mathbf{B} \right] - \left(\mathbf{v} \cdot \mathbf{B}\right)\left(\nabla \cdot \mathbf{B}\right)\,.
\end{equation}
From Eq.~(\ref{partialE}) and Eq.~(\ref{partialEandB1}), we obtain the decomposed energy conservation equation:
\begin{equation}\label{partialE1andB1}
\begin{aligned}
\frac{\partial E_1}{\partial t} &+ \nabla \cdot \left[ \left( E_1 + p \right) \mathbf{v} \right] = -\mathbf{v} \cdot \nabla \left(\mathbf{B}  \cdot \mathbf{B}\right)+\left(\mathbf{v} \cdot \nabla \mathbf{B}\right) \cdot \mathbf{B} \\&+\mathbf{B} \cdot \nabla \left(\mathbf{B} \cdot \mathbf{v}\right)-\left(\mathbf{B} \cdot \nabla \mathbf{v}\right) \cdot \mathbf{B}\,.
\end{aligned}
\end{equation}
More specifically, the terms $\mathbf{v} \cdot\nabla \left(\mathbf{B} \cdot \mathbf{B}\right)$ and $\mathbf{B} \cdot \nabla \left(\mathbf{B} \cdot \mathbf{v}\right)$ in Eq.~(\ref{partialE1andB1}) are commutated as:
$$
\begin{aligned}
\mathbf{v} \cdot\nabla \left(\mathbf{B} \cdot \mathbf{B}\right)&=  2~\left(\mathbf{v}\cdot\nabla\mathbf{B}\right)\cdot\mathbf{B},\\
\hbox{\rm and}\qquad\qquad\qquad\\
\mathbf{B} \cdot \nabla \left(\mathbf{B} \cdot \mathbf{v}\right)&=  \left(\mathbf{B}\cdot\nabla\mathbf{B}\right)\cdot\mathbf{v}+\left(\mathbf{B}\cdot\nabla\mathbf{v}\right)\cdot\mathbf{B}\,.
\end{aligned}
$$
Consequently, Eq.~(\ref{partialE1andB1}) is equivalent to:
\begin{equation}\label{partialE1andB1Simplied}
\frac{\partial E_1}{\partial t} + \nabla \cdot \left[ \left( E_1 + p \right) \mathbf{v} \right] =-~\mathbf{B}\cdot\left(\mathbf{v}\cdot\nabla\mathbf{B}\right)+\mathbf{v}\cdot\left(\mathbf{B}\cdot\nabla\mathbf{B}\right)
\,.
\end{equation}

Together with the mass conservation equation, momentum conservation equation, induction equation, and the HGLM, we obtain the following MHD equations with decomposed energy:
\begin{equation}\label{ExtendedDecomposision}
\left\{
\begin{array}{l}
\begin{aligned}
&\frac{\partial \rho}{\partial t} + \nabla \cdot (\rho \, \mathbf{v}) = 0, \\
&\frac{\partial (\rho \, \mathbf{v})}{\partial t} + \nabla \cdot \left[ \rho \mathbf{v \, v} + \left( p + \frac{\mathbf{B}^2}{2} \right) \mathbf{I} - \mathbf{B \, B} \right] =  \mathbf{0},  \\
&\frac{\partial E_1}{\partial t} + \nabla \cdot \left[ \left( E_1 + p \right) \mathbf{v} \right] = -~\mathbf{B}\cdot\left(\mathbf{v}\cdot\nabla\mathbf{B}\right)+\mathbf{v}\cdot\left(\mathbf{B}\cdot\nabla\mathbf{B}\right),\\
&\frac{\partial \mathbf{B}}{\partial t} + \nabla \cdot (\mathbf{v \, B} - \mathbf{B \, v} + \psi \mathbf{I}) = \mathbf{0},\\
&\frac{\partial \psi}{\partial t} + V_{\rm ref}^2~\nabla \cdot \mathbf{B} = 0\,.
\end{aligned}
\end{array} 
\right.
\end{equation}
For the coronal simulations, the source term includes extra terms, with gravity, heating and radiation loss defined as in \cite{Wang2025_FirsttimeevolvingCOCONUT,wang2025COCONUTMayEvent}, and thermal conduction defined as in \cite{wang2025COCONUTMayEvent,wang2025sipifvmtimeevolvingcoronalmodel}. For the 2D benchmark tests, no additional source terms are included in Eq.~(\ref{ExtendedDecomposision}).

Given that $\nabla\cdot\mathbf{B}=0$, the decomposed energy equation described in Eqs.~(\ref{partialE1andB1Simplied}) and (\ref{ExtendedDecomposision}) is identical to the  following equation:
\begin{equation}\label{partialE1inGAMARA}
\begin{aligned}
\frac{\partial E_1}{\partial t} + \nabla \cdot \left[ \left( E_1 + p \right) \mathbf{v} \right] &= -\mathbf{v} \cdot \left(\nabla \cdot \left(\frac{\mathbf{B}^2}{2} ~\mathbf{I}-\mathbf{B}~\mathbf{B} \right)\right)\\
&=\mathbf{v}\cdot\left(\nabla\times\mathbf{B}\times\mathbf{B}\right)+\left(\nabla\cdot\mathbf{B}\right)\left(\mathbf{v}\cdot\mathbf{B}\right)
\,.
\end{aligned}
\end{equation}  
In fact, the GAMERA (Grid Agnostic MHD for Extended Research Applications) code \citep{Zhang_2019}, an explicit MHD solver rebuilt from the LFM (Lyon–Fedder–Mobarry) MHD code \citep{LYON20041333} and widely used for simulations of the inner heliosphere, as well as Earth and planetary magnetospheres \citep{Merkin2016,Provornikova2024}, also adopts Eq.~(\ref{partialE1inGAMARA}) to solve the energy conservation equation.
In this paper, we further extend the energy decomposition approach to a fully implicit, full (thermodynamic) MHD coronal model to improve its numerical stability in simulating complex coronal evolutions during solar maximum CRs involving low $\beta$ near active regions.

Because we adopt an unstructured finite volume solver (COCONUT) and an unstructured geodesic mesh with prismatic elements \citep[i.e. with five planar faces;][]{Brchnelova2022,Perri_2022}, it is mandatory for us to perform the calculations in Cartesian coordinates. Accordingly, all calculations in this paper are carried out in a Cartesian coordinate system, and the formulations are presented in Cartesian form. For descriptions of the corresponding terms in the spherical coordinate system, we refer the reader to \citet{Hayashi2013}, \citet{Feng_2017}, and \cite{Feng2020book}.

\subsection{HLL solver with appropriately added dissipation term}\label{HLLhybridisedwithLLF}
In this paper, the inviscid flux $\mathbf{F}_n\left(\mathbf{U}_{i,L},\mathbf{U}_{j,R}\right)$ across the interface $\Gamma_{ij}$ of cells $i$ and $j$ is calculated using the approximate HLL-type Riemann solver \citep{Feng2020book,Wang_2022}. To improve numerical stability without obviously affecting the accuracy, we appropriately add an extra dissipation term to the HLL Riemann solver for the energy equation component in the low-$\beta$ region and describe it in the following form:
\begin{equation}\label{numericalflux}
\begin{aligned}
\mathbf{F}_{n}\left(\mathbf{U}_{i,L},\mathbf{U}_{j,R}\right)=&\frac{\mathbf{F}_n\left(\mathbf{U}_{i,L}\right)+\mathbf{F}_n\left(\mathbf{U}_{j,R}\right)}{2}-\frac{1}{2}\mathbf{D}^{'}_{\rm HLL}\left(\mathbf{U}_{i,L},\mathbf{U}_{j,R}\right)\,.
\end{aligned} 
\end{equation}
Here, $\mathbf{F}_n\left(\mathbf{U}\right)=\mathbf{F}\left(\mathbf{U}\right) \cdot \mathbf{n}_{ij}$ with $\mathbf{n}_{ij}$ denoting the unit normal vector of $\Gamma_{ij}$, and $\mathbf{U}_{i,L}$ and $\mathbf{U}_{j,R}$ denote the state variables reconstructed at the centroid of the interface $\Gamma_{ij}$ from cells $i$ and $j$, respectively,  using a second-order accurate least-squares reconstruction method \citep{BARTH1993,Lani2008AnOO} as described in \citet{Wang2025_FirsttimeevolvingCOCONUT}. As usual, a piecewise linear polynomial is used to reconstruct the primitive variables \citep{Wang2025_FirsttimeevolvingCOCONUT}. The difference here is that we do not apply limiters to the velocity and magnetic field to avoid accuracy degradation, while the Venkatakrishnan limiter \citep{VENKATAKRISHNAN1993} is retained for the remaining variables to suppress spatial oscillations.
In addition, $\mathbf{D}_{\rm HLL}^{'}\left(\mathbf{U}_{i,L},\mathbf{U}_{j,R}\right)$ represents the dissipation term of the numerical inviscid flux solver. 

For the HLL Riemann solver, the dissipation term is described as:
\begin{equation}\label{InviscidFluxHLL}
\begin{aligned}
\mathbf{D}_{\rm HLL}\left(\mathbf{U}_{i,L},\mathbf{U}_{j,R}\right)=&\frac{\left(S_L+S_R\right)\left(\mathbf{F}_{n}\left(\mathbf{U}_{j,R}\right)-\mathbf{F}_{n}\left(\mathbf{U}_{i,L}\right)\right)}{S_R-S_L}\\
&-\frac{2S_R~S_L}{S_R-S_L}\left(\mathbf{U}_{j,R}-\mathbf{U}_{i,L}\right).
\end{aligned}
\end{equation}
For local Lax-Friedrichs (LLF) solver \citep{Harten1983}, the dissipation term is described as:
\begin{equation}\label{InviscidFluxLLF}
\begin{aligned}
\mathbf{D}_{\rm LLF}\left(\mathbf{U}_{i,L},\mathbf{U}_{j,R}\right)=\max\left(|S_R|,~|S_L|\right)\left(\mathbf{U}_{j,R}-\mathbf{U}_{i,L}\right).
\end{aligned}
\end{equation}
The same as in \cite{Wang2025_FirsttimeevolvingCOCONUT}, $S_L$ and $S_R$ denote the minimum and maximum wave speeds, including the two waves with velocities $\pm V_{\rm ref}$ introduced by the HGLM method \citep{Dedner2002JCoPh}, 
and are defined as following:

\begin{equation}\label{SlandSr}
\begin{aligned}
S_{L}=&\min \left(0, ~\lambda_{\min}\left(\mathbf{U}_{i,L}\right),~\lambda_{\min}\left(\mathbf{U}_{j,R}\right),~-V_{\rm ref}\right)\\
S_{R}=&\max \left(0, ~\lambda_{\max}\left(\mathbf{U}_{i,L}\right),~\lambda_{\max}\left(\mathbf{U}_{j,R}\right),~V_{\rm ref}\right)
\end{aligned}\,,
\end{equation} 
where $$\lambda_{\min}\left(\mathbf{U}\right)=\min\left(V_n\left(\mathbf{U}\right),V_n\left(\mathbf{U}\right)\pm c_{f~{\rm or}~s}\left(\mathbf{U}\right),V_n\left(\mathbf{U}\right)\pm c_A\left(\mathbf{U}\right)\right)\,,$$$$\lambda_{\max}\left(\mathbf{U}\right)=\max\left(V_n\left(\mathbf{U}\right),V_n\left(\mathbf{U}\right)\pm c_{f~{\rm or}~s}\left(\mathbf{U}\right),V_n\left(\mathbf{U}\right)\pm c_A\left(\mathbf{U}\right)\right)\,,$$
with $c_{f~{\rm or}~s}\left(\mathbf{U}\right)=\sqrt{0.5\left(\frac{\gamma p}{\rho}+\frac{\mathbf{B}^2}{\rho} \pm \sqrt{\left(\frac{\gamma p}{\rho}+\frac{\mathbf{B}^2}{\rho}\right)^2-4\frac{\gamma p}{\rho}\frac{B_{n}^2}{\rho}}\right)}$ and $c_A\left(\mathbf{U}\right)=\frac{\left|B_{n}\right|}{\rho^{0.5}}$. 
Here, $V_n(\mathbf{U}) = \mathbf{v} \cdot \mathbf{n}_{ij}$ and 
$B_n(\mathbf{B}) = \mathbf{B} \cdot \mathbf{n}_{ij}$ denote the normal components of the velocity and magnetic field at the cell interface $\Gamma_{ij}$.

Given that MHD codes often suffer from non-physical pressure derived from the energy equation, and that the energy equation in Eq. (\ref{ExtendedDecomposision}) is no longer in conservative form, making the discretization error of the right-hand-side terms potentially a source of the instability, we improve the numerical stability by appropriately incorporating a portion of the LLF dissipation term into the HLL dissipation when computing the energy equation component of the inviscid flux, while retaining the original HLL solver for the other components to preserve higher accuracy.
We have 
\begin{equation}\label{InviscidFluxLLFandHLL}
\begin{aligned}
\mathbf{D}_{\rm HLL}^{'}\left(\mathbf{U}_{i,L},\mathbf{U}_{j,R}\right)=&\mathbf{D}_{\rm HLL}\left(\mathbf{U}_{i,L},\mathbf{U}_{j,R}\right)\\&+\alpha \mathbf{{\cal{R}}}~\mathbf{D}_{\rm LLF}\left(\mathbf{U}_{i,L},\mathbf{U}_{j,R}\right)\,,
\end{aligned} 
\end{equation}
where 
$\mathbf{{\cal{R}}}=\begin{pmatrix} \mathbf{0}_{7\times 7} \quad 0\quad 0\\ 0 \quad 1 \quad 0 \\ 0 \quad 0 \quad 0  \end{pmatrix} \,
$, with $\mathbf{0}_{7\times 7}$ being the zero matrix of $7 \times 7$. 
$\alpha=\kappa \cdot \max\left(\tanh\left(0.01\big/\beta_i\right),\tanh\left(0.01\big/\beta_j\right)\right)$ is an adjustable parameter used to adjust the amount of the added LLF dissipation term, with $\beta_i$ and $\beta_j$ being the $\beta$ calculated at the centroids of cells $i$ and $j$, respectively, and $\kappa=0.125$ is adopted in this paper.

\subsection{Implementation of boundary conditions}\label{BCinner}
In this paper, we first validate the decomposed energy strategy by comparison with the full energy equation approach using quasi-steady-state coronal simulations during the solar minimum and rising phases. These simulations are constrained by a single static GONG-ADAPT (Global Oscillation Network Group-Air Force Data Assimilative Photospheric Flux Transport) synoptic magnetogram \citep{Arge2010,Hickmann_2015}\footnote{\url{https://gong.nso.edu/adapt/maps/gong/2011/}} from August~3,~2008, and September~10,~2021, which represents the solar minimum and rising phases of the solar cycle, respectively. We also validate this energy decomposition strategy in a solar maximum time-evolving coronal simulation. A quasi-steady-state coronal simulation constrained by a static magnetogram on March 29, 2025, the beginning of Carrington Rotation (CR) 2296, provides the initial solution for the subsequent time-evolving simulation driven by about 660 hourly updated GONG-zqs (zero-point-corrected) photospheric magnetograms\footnote{\url{https://gong.nso.edu/data/magmap/QR/zqs/202405/}} \citep{Hill2018,LiHuichao2021,Perri_2023}.
These simulations are performed in a quasi-inertial coordinate system, with the Earth permanently positioned at $\phi=60^{\circ}$.
The radial magnetic field $B_r$ in the synoptic GONG-zqs magnetograms is provided in a heliocentric inertial (HCI) coordinate system \citep{Burlaga1984MHDPI,FRANZ2002217} and the time interval between two temporally adjacent files during this period is approximately one hour with fluctuations of several minutes. For ease of implementation, we first rotate the original magnetograms into a heliographic co-rotating coordinate system and apply cubic Hermite interpolation to generate a sequence of magnetograms with an exact one-hour cadence. These exactly hourly-updated magnetograms are then rotated back into the HCI coordinate system to drive the time-evolving coronal simulations. Meanwhile, the inner-boundary magnetic fields at each physical time step are interpolated from the four temporally nearest hourly updated input magnetograms by applying a cubic Hermite interpolation \citep{Wang2025_FirsttimeevolvingCOCONUT}. 

Considering that the original magnetograms exhibit steep gradients in the radial magnetic field, which can lead to numerical divergence \citep{Hayashi2013}, we employ a PF solver with the 0th-order and all spherical harmonic components above the 25th order removed \citep{Toth_2011,Perri_2022}  to extrapolate the photospheric magnetograms, where magnetic field strengths can reach several hundred Gauss near active regions, to the bottom of the low corona \citep{Perri_2022,Perri_2023,Kuzma_2023}.
Unlike the simulations in \citet{wang2025COCONUTMayEvent}, which used a 20th-order spherical harmonics PF solver with a filter to smooth the magnetic field, limiting its strength in the low corona to not significantly exceed 30 Gauss, we directly use a 25th-order PF solver without additional filter, allowing magnetic field strengths to exceed 100 Gauss near the active regions in the low corona during the time-evolving simulation. To minimize the influence of the preprocessing on the global magnetic-field structure, the reconstructed magnetic fields are evaluated at the same pixel locations as the original magnetograms on the surface at 1~$R_s$. The filtered magnetograms are used to define the radial magnetic field at the inner boundary. Meanwhile, the inner-boundary velocity and tangential magnetic field are specified following \citet{Wang2025_FirsttimeevolvingCOCONUT}, and the density and pressure are set as in \citet{wang2025COCONUTMayEvent}.
 
For simulations of the solar maximum and increasing phases, a standard set of parameters is adopted, the temperature and plasma density at the inner boundary are set to $1.8 \times 10^6~\rm K$ and $3.34 \times 10^{-13}~\rm kg~m^{-3}$, respectively, and the thermal pressure at the inner boundary is specified as $0.01~\rm Pa$ \citep{wang2025COCONUTMayEvent,Baratashvili2025}. 
Due to the limitations of the currently adopted, overly simplified heating source terms, which are proportional to the magnetic field strength, the heating and acceleration effects are much more pronounced during solar maximum (when the magnetic field is significantly stronger) than during solar minimum. We noticed that higher or lower plasma densities can mitigate the excessively high or insufficiently low radial velocities during solar maximum or minimum, respectively. Therefore, we adopted different plasma parameters for solar maximum and minimum conditions.
For the solar minimum simulation, the thermal pressure and plasma density at the inner boundary are set to $0.0042~\rm Pa$ and $1.67 \times 10^{-13}~\rm kg~m^{-3}$, respectively, with the corresponding temperature defined as $1.5 \times 10^6~\rm K$ \citep{Baratashvili,Wang2025_FirsttimeevolvingCOCONUT}. Additionally, the velocity vector, the tangential magnetic field at the inner boundary, and the outer boundary conditions are treated the same as in \cite{Wang2025_FirsttimeevolvingCOCONUT}. 

\section{Numerical results}\label{sec:Numerical Results}
The decomposed energy strategy is validated in both quasi-steady-state and time-evolving coronal simulations, as summarized in Table~\ref{tab:R1}, together with two 2D test cases.
\begin{table}
\centering
\caption{Summary of coronal simulations using full and decomposed energy approaches}
\label{tab:R1}
\begin{tabular}{lllll}
\hline\noalign{\smallskip}
CR No. & Solar cycle phase & Modeling regime & Energy equation & Performance \\
\noalign{\smallskip}\hline\noalign{\smallskip}
2073 & minimum & static & full & stable (no crash) \\
2073 & minimum & static & decomposed & nearly identical results \\
2248 & rising  & static & full & stable (no crash) \\
2248 & rising  & static & decomposed & locally different results \\
2296 & maximum & static & full & unstable (crash) \\
2296 & maximum & static & decomposed & stable (no crash) \\
2296 & maximum & time-evolving & full & unstable (crash) \\
2296 & maximum & time-evolving & decomposed & stable (no crash) \\
\noalign{\smallskip}\hline
\end{tabular}
\end{table} 
In this section, we first perform quasi-steady-state coronal simulations using both versions of COCONUT: one that adopts the full energy equation \citep{wang2025COCONUTMayEvent} and the other that uses the decomposed energy strategy described in this paper, to compare the results of the energy decomposition method with those of the traditional full energy approach. As described in Subsection \ref{BCinner}, these simulations are constrained by static magnetograms from CR 2073 and CR 2248, respectively. CR 2073 corresponds to the period from August 3 to August 30, 2008, which falls within Solar Cycle (SC) 23 and is near the solar minimum of this cycle. CR 2248 spans from August 28 to September 24, 2021 and falls within SC 25 during the early rising phase of the cycle. For the solar minimum and increasing cases, the maximum magnetic field strength near the solar surface is approximately 8.5 and 42.1 Gauss, respectively. In order to show the suitability of our numerical approach to tackle these challenging coronal simulations, Appendix \ref{sec:benchmark} presents some verification results for two 2D test-cases which are often used to benchmark MHD solvers: the Orszag–Tang vortex and the MHD rotor. These simulations allow for evaluating the effects of the decomposed energy equation and the additional dissipation term in the energy component of the HLL Riemann solver, demonstrating the ability of our scheme to preserve accuracy (with respect to reference solutions) in relatively complex flows.

After validating the coronal model COCONUT that adopts the decomposed energy equation through the above comparison, we further perform a time-evolving coronal simulation using this version of COCONUT to demonstrate its capability to capture solar maximum coronal evolutions with complex and strong magnetic fields.
This time-evolving simulation starts from a quasi-steady-state coronal simulation result and then is driven by the subsequent hourly-updated magnetograms during CR 2296, which spans from March 29 to April 26, 2025 and falls within the solar maximum phase of Solar Cycle 25. During the simulation, the maximum magnetic field strength near the active regions in the low corona generally exceeds 55 Gauss and occasionally reaches up to 100 Gauss. Executed on 360 CPUs of the wICE cluster, part of the Tier-2 supercomputer at the Vlaams Supercomputer Centrum\footnote{\url{https://www.vscentrum.be/}}, the simulation runs 25 times faster than real-time. 

\subsection{Decomposed versus full energy equation approach in quasi-steady-state simulation results}\label{sec:DecEversusFullE}
\begin{figure}[htpb]
\begin{center}
  \vspace*{0.01\textwidth}
    \includegraphics[width=0.9\linewidth,trim=1 1 1 1, clip]{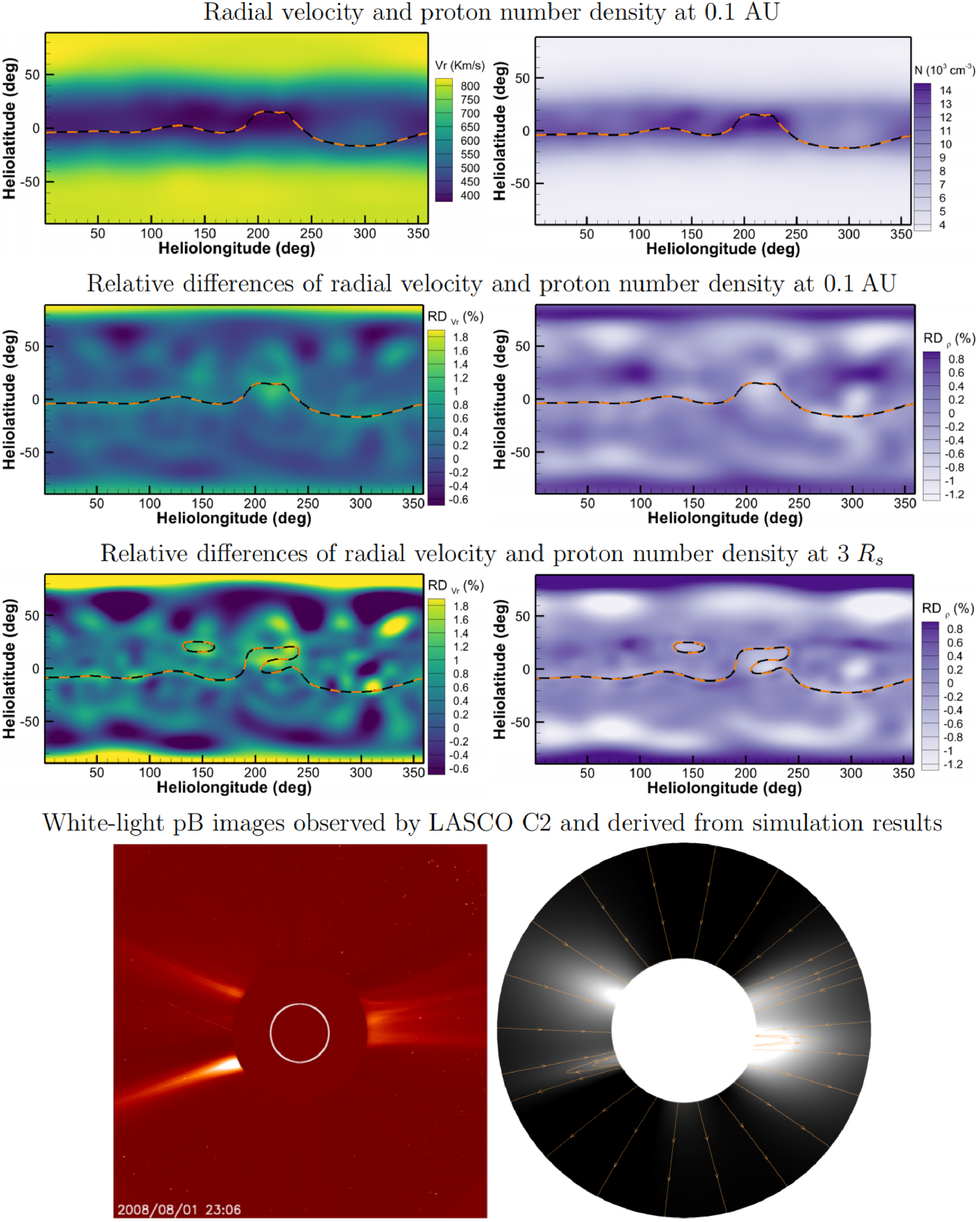}
\end{center}
\caption{Distribution of the radial plasma speed $V_r \, (\mathrm{km~s^{-1}})$ and proton number density $N \, (10^3~\mathrm{cm^{-3}})$ at 0.1 AU calculated using COCONUT with the decomposed energy equation (first row). The relative differences in radial velocity and plasma density  between the results obtained with the decomposed and the traditional full-energy equations at 0.1 AU (second row) and 3~$R_s$ (third row) are also shown. The fourth row presents the white-light pB images observed by LASCO/SOHO and synthesized from electron density which is treated to be the same as the proton density simulated by COCONUT, spanning a range from 2.3 to 6$\;R_s$ with the decomposed energy equation. The black solid and orange dashed lines represent the MNLs calculated using the decomposed and full energy equations, respectively. The orange solid lines denote the magnetic field lines calculated with the decomposed energy equation. These figures correspond to CR 2073.} 
\label{localdiffCR2073}
\end{figure}
Figure~\ref{localdiffCR2073} presents the simulated radial velocity ($V_r$ in unite of $\mathrm{km~s^{-1}})$ and proton number density (N in unite of $10^3~\mathrm{cm^{-3}}$; first row) for CR 2073 at 0.1 AU, as well as the relative differences in radial velocity $\rm RD_{V_r}=\left(V_r^{FullE}-V_r^{DecE}\right)\big/V_r^{FullE}$  and plasma density $\rm RD_{\rho}=\left(\rho^{FullE}-\rho^{DecE}\right)\big/\rho^{FullE}$ at 0.1~AU (second row) and 3~$R_s$ (third row). The superscripts ``$^{\rm DecE}$" and ``$^{\rm FullE}$" denote the variables calculated by COCONUT with the decomposed and full energy equations, respectively. 
The bottom-left panel show Level~2 white-light polarized brightness (pB) image of the solar corona observed by the C2 coronagraph of the three-telescope LASCO (Large Angle and Spectrometric Coronagraph) instrument \citep{Brueckner1995,Frazin2012} onboard the ESA/NASA Solar and Heliospheric Observatory  spacecraft\footnote{\url{https://stereo-ssc.nascom.nasa.gov/browse/}}. In these observations, direct sunlight is blocked by an occulting disk to create an artificial eclipse. The white circle in the images denotes the position of the solar disk. The bottom-right panel present the white-light pB image synthesized by line-of-sight (LoS) integration of the radiant intensity generated by Thomson scattering of photospheric light by coronal electrons \citep{inhester2016}. The pB for a given pixel position on the selected plane-of-the-sky is calculated as
\begin{equation*}
    pB = \int_{\mathrm{LoS}} n_e(s)\,\bigl(I_{\mathrm{tan}} - I_{\mathrm{rad}}\bigr)\, \mathrm{d}s ,
\end{equation*}
where $s$ and $n_e(s)$ are the distance and electric number density (assumed identical to proton number density) along the LoS, $I_{\mathrm{tan}}$ and $I_{\mathrm{rad}}$ denote the tangential and radial polarization components of the radiant intensity scattered from a single electron, respectively, and are calculated the same as in \citet{AsensioRamos2023} and \citet{Sorokina2025}. Meanwhile, a Fourier-normalizing radial gradient filter \citep{DruckmUllerovA_2011} is applied to enhance the coronal structures and suppress the radial brightness gradient.
The results show that the simulations from both approaches are nearly identical, with relative differences in radial velocity and plasma density of less than $1\%$ in most regions. The magnetic neutral line (MNL) obtained from the two methods also aligns closely. This comparison of a solar minimum CR shows that the COCONUT with the decomposed energy equation is fully consistent with the COCONUT that uses the full energy equation. Moreover, the simulated white-light pB image successfully reproduces the two bright structures observed on the east limb by the LASCO C2. The bright structure on the west limb, located around the solar equator, is also captured in the simulated pB image.
  
\begin{figure}[htpb]
\begin{center}
  \vspace*{0.01\textwidth}
    \includegraphics[width=0.9\linewidth,trim=1 1 1 1, clip]{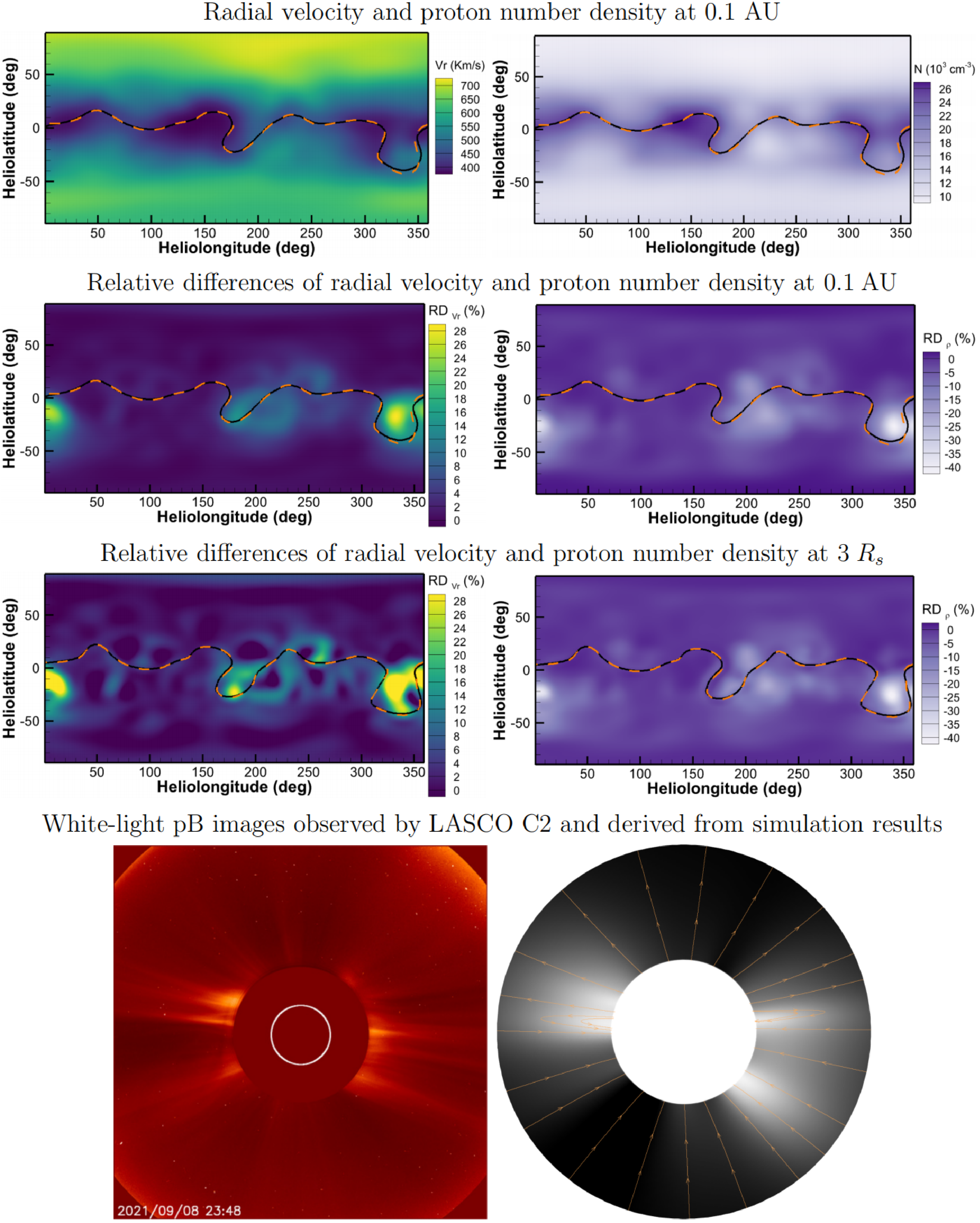}
\end{center}
\caption{Same as Figure~\ref{localdiffCR2073}, but for CR 2248.}\label{localdiffCR2248}
\end{figure}
After validating the reliability of COCONUT with the decomposed energy equation by comparing its simulation results with those from the traditional full energy equation in a quasi-steady-state coronal simulation of a solar minimum CR, we further extended the comparison to a CR during the rising phase of solar activity, characterized by stronger and more complex magnetic fields, with the field near active region in the low corona reaching up to 40 Gauss.
Figure~\ref{localdiffCR2248} shows that the MNL is consistent between the simulations using the decomposed and full energy equations, with only minor deviations around the ``U"-shaped segment between $320^{\circ}$ and $360^{\circ}$ in longitude. In addition, the simulated high-density, low-speed (HDLS) regions are mainly concentrated around the MNL, indicating that the simulated HDLS streams coincide with the current sheets \citep{Wang_2007,Decraemer_2019}. The relative differences in radial velocity ($\rm RD_{V_r}$) and plasma density ($\rm RD_{\rho}$) at 0.1 AU range from $-0.55\%$ to $28.65\%$ and from $-41.38\%$ to $3.94\%$, respectively. At 3~$R_s$, $\rm RD_{V_r}$ and $\rm RD_{\rho}$ range from $-10.37\%$ to $44.27\%$ and from $-43.33\%$ to $1.82\%$, respectively. Moreover, $\left|\rm RD_{V_r}\right|$ and $\rm RD_{\rho}$ remain below $10\%$ and $15\%$ in most regions, while areas with $\rm RD_{V_r}>10\%$ or $\rm RD_{\rho}<-15\%$ are mainly concentrated near the ``U"-shaped segment of the MNL. 

Furthermore, we list the average relative differences in the solution variables between the results obtained using the decomposed and full energy equations in Table \ref{QSSVSTE}, denoted as ${\rm RD}_{{\rm ave},\chi}$:
$${\rm RD}_{{\rm ave},\chi}=\sum\limits_{i=1}^{N_{\rm cell}} \left(\frac{\left|\chi_i^{{\rm FullE}}-\chi_i^{\rm DecE}\right|}{\chi_{\rm ave}^{{\rm FullE}}} \cdot \omega_i\right)\,, $$
where $\omega_i=V_i\big/\sum\limits_{i=1}^{N_{\rm cell}}V_i$ is the volume weighting with $V_i$ denoting the volume of cell $i$, $\chi_i \in \{~ |B_r|_i,~ |V_r|_i,~\rho_i,~|\rho V_r|_i ~\}$ represents the corresponding variable at the centroid of cell $i$, $\chi_{\rm ave}^{{\rm FullE}}=\sum\limits_{i=1}^{N_{\rm cell}}\left(\chi_i^{\rm FullE}\cdot \omega_i\right)$, and $N_{\rm cell}$ is the total number of discretized cells in the computational domain.
This indicates that, for the solar minimum case, the overall differences in plasma density, radial velocity, mass flux, and radial magnetic field strength between the decomposed-energy and full-energy COCONUT simulations remain below $1\%$, confirming the consistency of the decomposed energy equation with the full energy equation. In contrast, for the increasing-phase case with a stronger and more complex magnetic field, the average relative differences increase by nearly an order of magnitude but remain below $4.5\%$ for all these four parameters. 
\begin{table}
\centering
\caption{Average relative differences between the results with the decomposed and full energy equations.}
\label{QSSVSTE}
\begin{tabular}{lllll}
\hline\noalign{\smallskip}
 Parameters & ${\rm RD}_{{\rm ave},\rho}$  & ${\rm RD}_{{\rm ave},\left|V_r\right|}$& ${\rm RD}_{{\rm ave},\left|\rho V_r\right|}$ & ${\rm RD}_{{\rm ave},\left|B_r\right|}$ \\
\noalign{\smallskip}\hline\noalign{\smallskip}
 $\rm CR~2073$ & $0.29\%$  & $0.26\%$ & $0.51\%$ & $0.52\%$ \\
 $\rm CR~2248$ & $4.02\%$  & $3.28\%$  & $2.51\%$ & $3.17\%$ \\
\noalign{\smallskip}\hline
\end{tabular}
\end{table}

Together with Figures~\ref{localdiffCR2073} and ~\ref{localdiffCR2248}, and Table \ref{QSSVSTE}, these results show that the decomposed energy equation yields nearly identical results when the magnetic field is relatively simple and weak, as is typical during solar minimum, while producing noticeable differences in certain regions with strong and complex magnetic field configurations, as commonly seen during the increasing phase, yet still maintaining overall consistency with the full energy equation. Additionally, the COCONUT version using the full energy equation crashed during quasi–steady-state coronal simulation for CR 2296, a solar maximum case with an even stronger and more complex magnetic field than the increasing phase case of CR 2248. In contrast, the COCONUT using the decomposed energy equation performed well, further demonstrating its numerical stability advantages in simulating coronal structures with strong and complex magnetic fields. 

\begin{figure}[htpb]
\begin{center}
  \vspace*{0.01\textwidth}
    \includegraphics[width=0.9\linewidth,trim=1 1 1 1, clip]{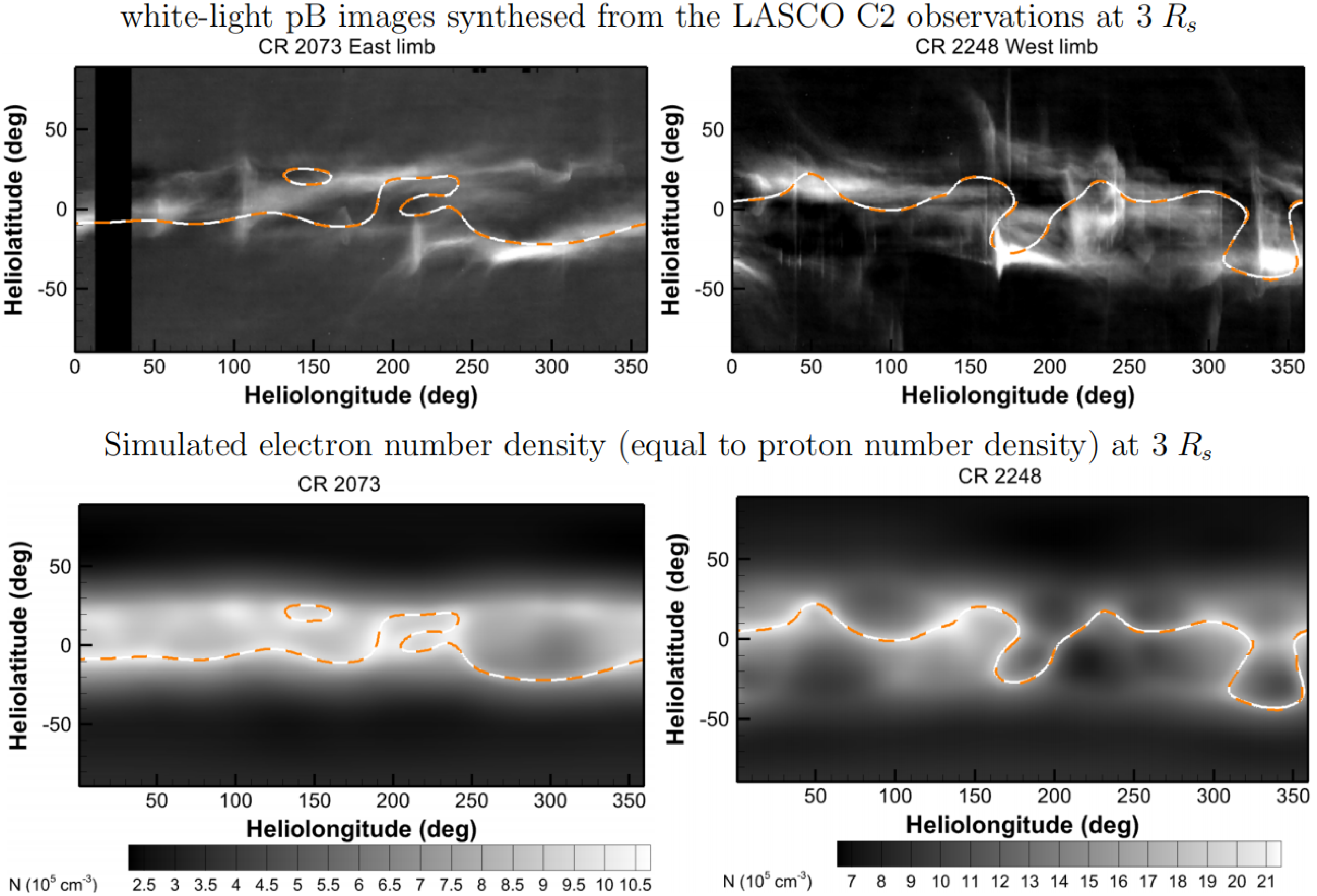}
\end{center}
\caption{Synoptic maps of east (top left) and west limb (top right) white-light pB observations from the  LASCO C2 instrument onboard SOHO spacecraft at 3 $R_s$ for CRs 2073 and 2248, alongside the simulated electron number density $N \, (10^5~\mathrm{cm^{-3}})$ which is treated to be identical to the proton density} (bottom) calculated with the decomposed energy equation. The white solid and orange dashed lines represent the MNLs derived from the simulation results calculated using the decomposed and the full energy equations, respectively.\label{pBat3Rs}
\end{figure}
Figure~\ref{pBat3Rs} presents the white-light pB images at $3~R_s$, synthesized from east- and west-limb observations by the Large Angle and Spectrometric Coronagraph C2 (LASCO-C2; \citealt{Brueckner1995}) onboard the Solar and Heliospheric Observatory (SOHO)\footnote{\url{http://lasco-www.nrl.navy.mil/carr_maps/c2/}}, for CRs 2073 (top left) and 2248 (top right), together with the simulated electron number density, treated to be the same as the proton density, at $3~R_s$ (bottom) calculated using the decomposed energy equation. It can be seen that the simulated MNLs from the decomposed and full energy equations are nearly identical. For both CRs, the simulated MNLs align well with the central axis of the observed bright pB strip structures, although the observed bright segments deviate by up to about $20^{\circ}$, northward between $130^{\circ}$ and $190^{\circ}$ for CR 2073, and southward between $185^{\circ}$ and $210^{\circ}$ for CR 2248, relative to the simulated MNLs. Correspondingly, the bright structures in the simulated plasma density distribution, representing high-density, low-speed plasma flows that are usually concentrated around the MNLs, are consistent with those observed in the pB images. Incorporating more self-consistent coronal heating and solar wind acceleration mechanisms, rather than the empirical treatments adopted in this paper, may further improve the consistency between the simulated and observed results.

\begin{figure*}[htpb]
\begin{center}
  \vspace*{0.01\textwidth}
    \includegraphics[width=0.9\linewidth,trim=1 1 1 1, clip]{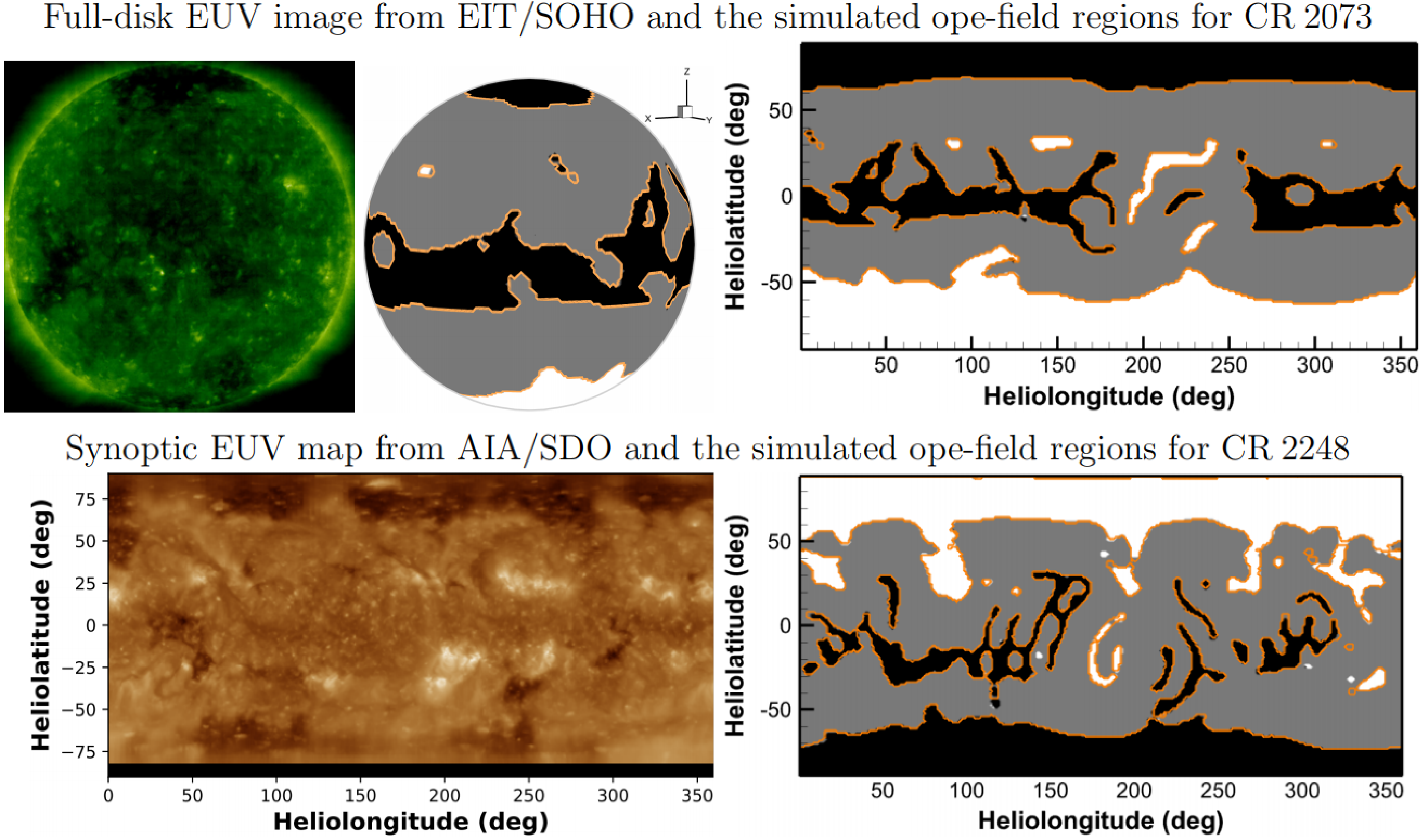}
\end{center}
\caption{Full-disk image from EIT/SOHO at 195~\AA\ on 2008 August 3 (top left), the corresponding simulated full-disk image of open- and closed-field regions obtained with the decomposed energy equation (top middle), and the global distributions of open- and closed-field regions derived from COCONUT simulations with the decomposed energy equation for CR 2073 (top right) and CR 2248 (bottom right). The white and black patches denote open-field regions, where magnetic field lines point outward and inward, respectively, relative to the Sun. The grey regions represent closed-field regions. The orange solid lines overlaid on these contours denote the edge of open-field regions derived from simulation results with the full energy equation. The bottom-left panel shows the synoptic EUV map from the 193~\AA\ channel of AIA/SDO for CR 2248.}\label{CHquasisteady}
\end{figure*}
In Figure~\ref{CHquasisteady}, we compare the distributions of open-field regions derived from both the decomposed and full energy equations, and contrast them with extreme ultraviolet (EUV) observations. 
Coronal holes (CHs) are well-established as the main contributors to high-speed solar wind \citep{Nolte1976,CRANMER20023,Hofmeister2022}, and are among the most prominent and dynamic structures of the solar corona \citep{Linker1999JGR,Hayes2001,Frazin2007,Petrie2011SoPh,FengMa2015,Feng_2017}. With their magnetic field lines open to interplanetary space, resulting in low plasma density, these regions typically appear as dark areas in EUV and soft X-ray images. Figure~\ref{CHquasisteady} shows that the edges of open-field regions derived from both the decomposed and the full energy equations well capture the observed polar CH regions observed in the synoptic EUV image, synthesized from the 193 {\AA} channel of Atmospheric Imaging Assembly (AIA) instrument \citep{Lemen2012} onboard the Solar Dynamics Observatory (SDO) spacecraft \citep{Pesnell_2012SoPh}\footnote{\url{https://sdo.gsfc.nasa.gov/data/synoptic/}}, from about $60^\circ$ latitude poleward in the north and from around $70^\circ$ latitude poleward in the south. 
The observed isolated CHs centered at 
$(\theta_{\rm lat}, \phi_{\rm long}) = (15^\circ \,\mathrm{S},\, 50^\circ)$ 
and $(15^\circ \,\mathrm{S},\, 300^\circ)$, the leading CHs located near 
longitudes $10^\circ$, $90^\circ$, $190^\circ$, $280^\circ$, and $330^\circ$, 
which extend from the northern pole to approximately $25^\circ \,\mathrm{N}$, 
and the leading CH spanning between longitudes $220^\circ$ and $250^\circ$, extending from 
the southern pole to about $25^\circ \,\mathrm{S}$, all have corresponding 
patches in the simulated open-field regions.
Additionally, the simulated isolated open-field regions between 
$25^\circ\,\mathrm{N}$ and $25^\circ\,\mathrm{S}$ correspond to relatively dark areas in the low-latitude portions of the observed synoptic EUV image. Since the synoptic EUV image for CR~2073 is unavailable, we present instead the full-disk EUV image observed by EIT/SOHO on August~3, 2008. This image shows that the simulated open-field regions successfully capture the observed polar CHs, as well as the isolated CHs distributed at low latitudes.

\subsection{Time-evolving solar maximum simulation by energy decomposition approach}\label{sec:DecEversusFullE}

\begin{figure*}[htpb]
\begin{center}
  \vspace*{0.01\textwidth}
    \includegraphics[width=0.8\linewidth,trim=1 1 1 1, clip]{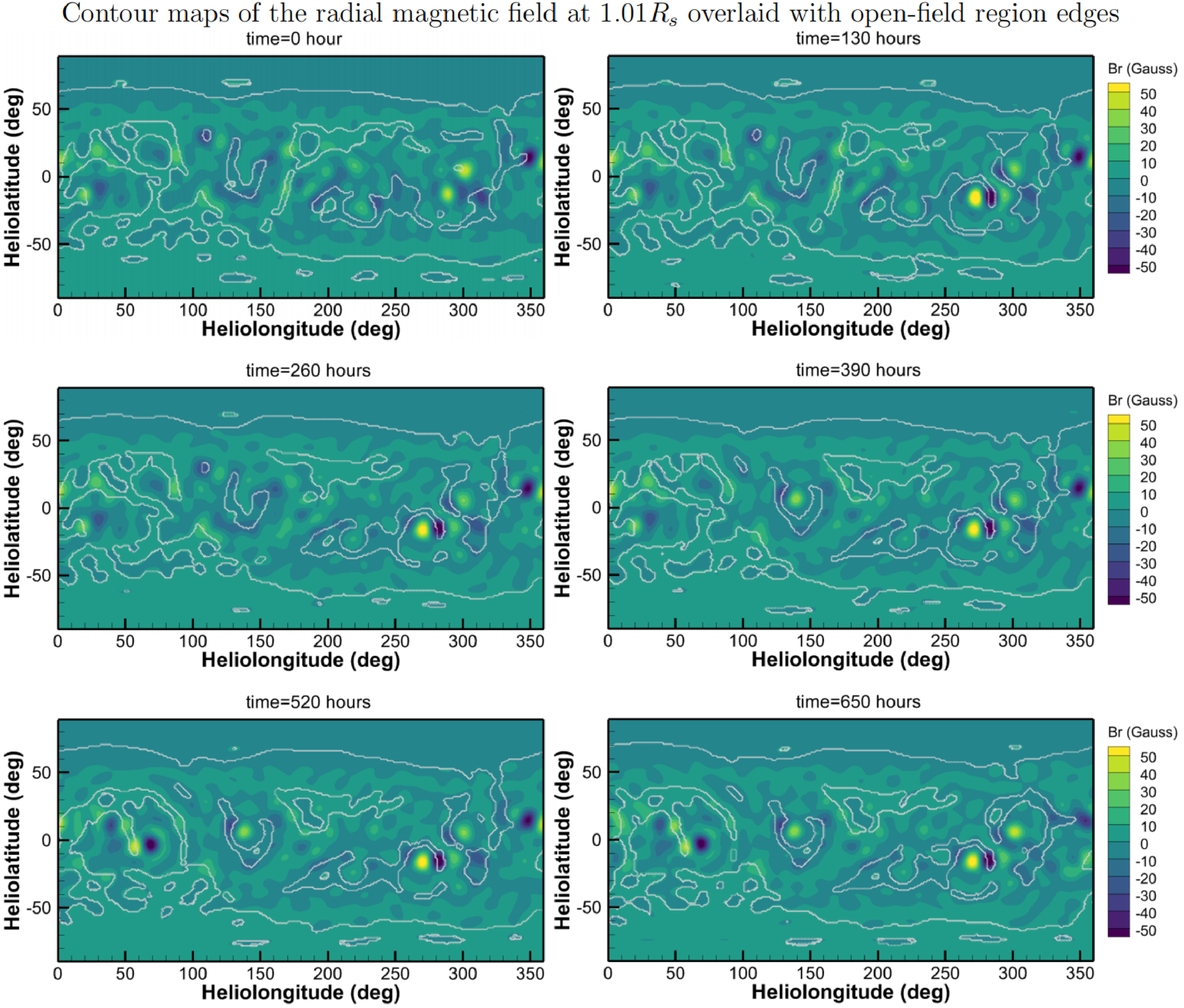}
\end{center}
\caption{Distributions of the inner-boundary radial magnetic field at the 0th, 130th, 260th, 390th, 520th, and 650th hours of the time-evolving coronal simulation during CR 2296. White lines overlaid on the magnetic field contours indicate the boundaries of the open-field regions derived from COCONUT simulation results using the decomposed energy equation.}\label{MagandCHevolution}
\end{figure*}
\begin{figure*}[htpb]
\begin{center}
  \vspace*{0.01\textwidth}
    \includegraphics[width=0.8\linewidth,trim=1 1 1 1, clip]{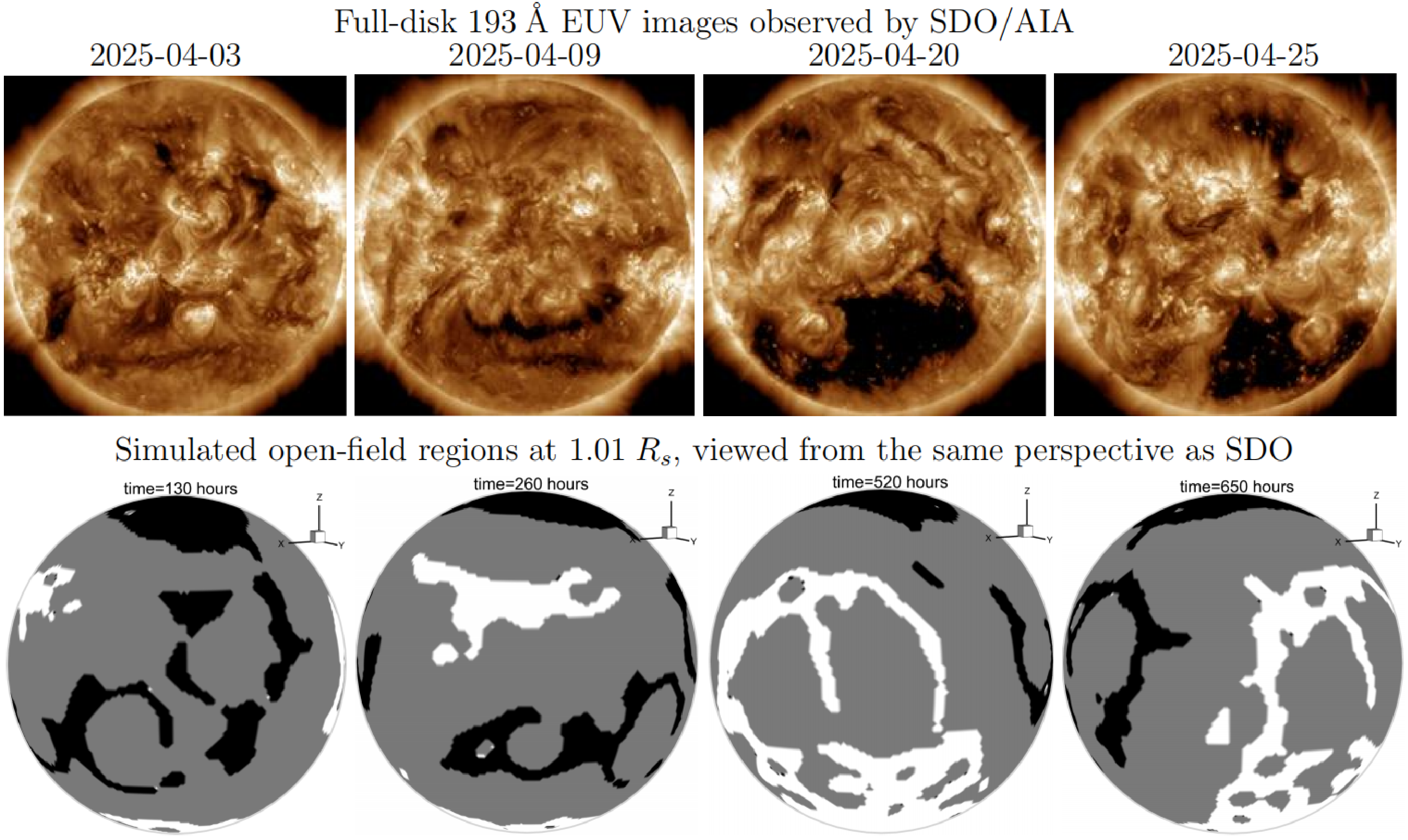}
\end{center}
\caption{Full-disk images from SDO/AIA at 193~\AA\ on April 3, 9, 20, and 25, 2025 (top), and the corresponding simulated full-disk distributions of open- and closed-field regions from the time-evolving COCONUT simulation for CR 2296 using the decomposed energy equation (bottom). The white and black patches denote open-field regions with magnetic field lines pointing outward and inward relative to the Sun, and the grey regions denote closed-field regions.}\label{CHFullDisk}
\end{figure*}

Figure~\ref{MagandCHevolution} displays six snapshots of the inner-boundary radial magnetic field distribution at different moments during the coronal evolution in CR~2296. These images are shown in the co-rotating coordinate system. The white solid lines overlaid on the radial magnetic field contours denote the boundaries of the open-field regions, obtained by tracing magnetic field lines calculated with the decomposed energy equation back to the inner boundary. It is evident that the distribution of open-field regions in this figure is far more complex than during the rising phase, and that the evolution of the open-field regions and the corresponding CHs (also shown in Figure~\ref{CHFullDisk}) during this solar maximum period is particularly notable. It is also observed that, for the dipole structure near the solar surface, the pole with stronger magnetic field strength is typically located within a closed-field region, whereas the opposite pole is often dominated by open-field regions. It is noted that at the 130th hour of the simulation, the emergence of a dipole 
centered at $\left(\theta_{\rm lat}, \phi_{\rm long}\right) = \left(15^\circ\,\mathrm{S},\,280^\circ\right)$ enlarges the open-field region surrounding the lower-left portion of the positive pole, 
where the magnetic field points outward from the Sun, and gives rise to a narrow strip extending across the negative pole. At the 520th hour, the dipole centered at $\left(\theta_{\rm lat}, \phi_{\rm long}\right) = \left(5^\circ\,\mathrm{S},\,65^\circ\right)$ enlarges 
the closed-field region around the negative pole, and leads to a narrow strip passing through the positive pole. By the 650th hour, the strengthening of the positive pole of the dipole centered at $\left(\theta_{\rm lat}, \phi_{\rm long}\right) = 
\left(10^\circ\,\mathrm{N},\,290^\circ\right)$ further enlarges the closed-field region around the positive pole.  Besides, it is noticed that some ring-like coronal-hole patterns are present in these simulation results. These features may be related to ringing artifacts associated with the preprocessing by the PF solver with limited order of spherical harmonics \citep{Toth_2011}. In future work, we will attempt to mitigate this effect by deriving the inner-boundary magnetic field for coronal simulations using a filtered spherical-harmonic PF solver \citep{MCCLARREN20105597, wang2025COCONUTMayEvent, Murteira2025} or a Poisson solver prescribed by the observed photospheric magnetograms \citep{Toth_2011,LiuXJ_2019}.

In Figure~\ref{CHFullDisk}, we further compare the simulated open-field regions with the 193~\AA\ EUV images observed by SDO/AIA. The results show that at the 130th hour in the simulation, the simulated open-field regions reproduce the observed dark areas, which are scattered along the lower and right edges of the solar disk. 
At the 260th hour, the simulated southern-hemisphere open-field region (dark patch) captures the observations, and the northern-hemisphere ``$\pi$"-shape open-field region (white) likewise captures the relative dark regions seen in the EUV image. At the 520th hour, the simulated ``M"-shape open-field region (white patch) captures the observation, although the simulated left ``leg" is about $10~^{\circ}$ eastern than that in the observed ``M"-shape dark region. The simulated open-field region, which covers the bottom of the southern pole, corresponds to the large CH near the southern pole in the observed EUV image. At the 650th hour, the simulated open-field region (white patch) at the west limb captures the observed dark regions in the EUV image. The ``n"-shape dark region at the east limb of the observed EUV image is also captured by the simulated open-field region (dark). There are also observed dark regions that are not captured in the simulated results, or vice versa. For example, the relatively dark line structure on the right-hand side of the ``$\pi$"-shape structure in the observation of April 9, 2025, is missing in the simulation result; the simulated open-field region covering the northern pole is also missing in the observations. The discrepancy may be attributed to the resolution limitations of both synoptic magnetograms and the grid mesh used in this simulation, as well as to a lack of reliable magnetic field observations in the high-latitude region. Limiting only the adopted radial magnetic field to the observed one may also lead to a discrepancy between the simulation and the observed results.

\begin{figure*}[htpb]
\begin{center}
  \vspace*{0.01\textwidth}
    \includegraphics[width=0.8\linewidth,trim=1 1 1 1, clip]{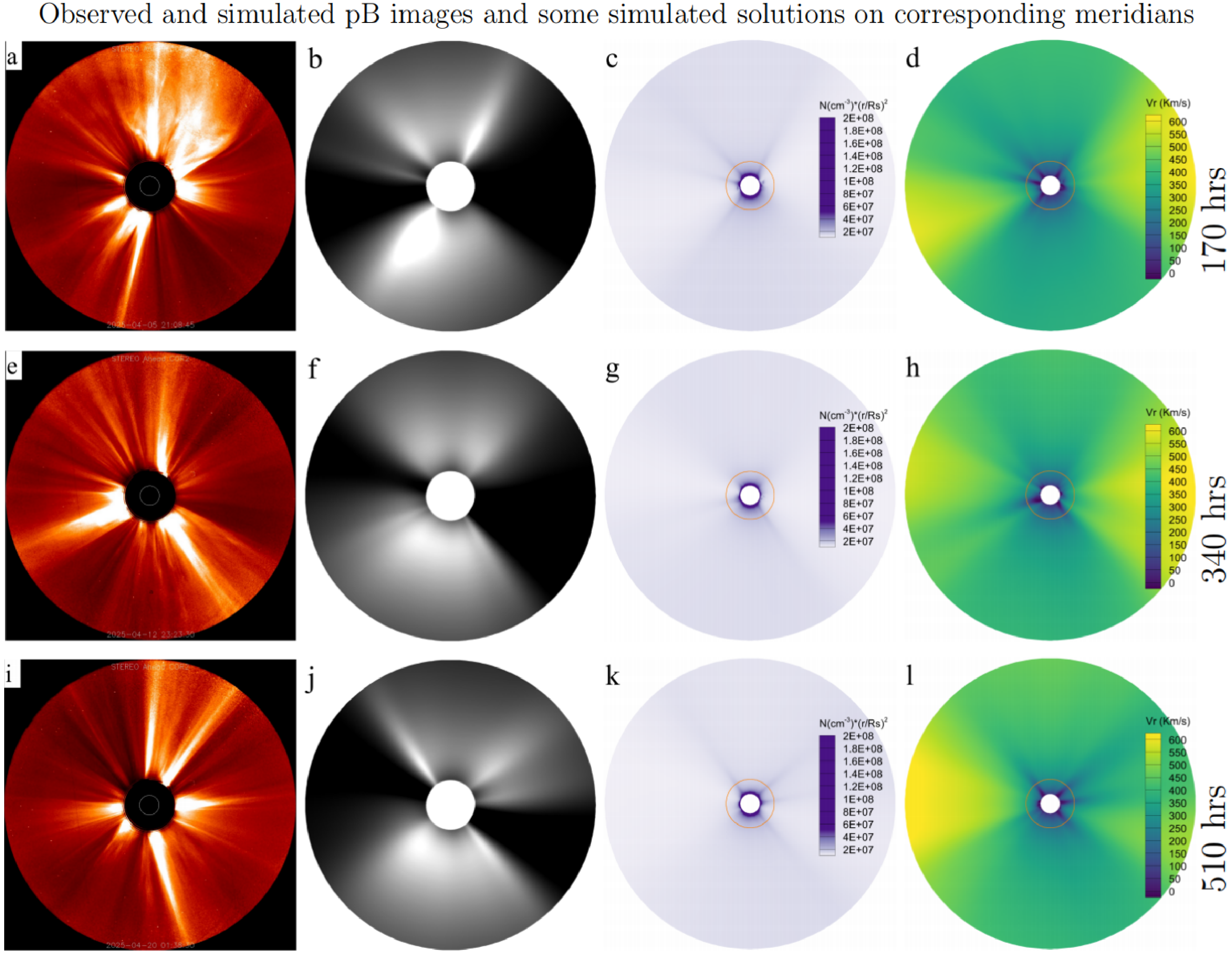}
\end{center}
\caption{White-light pB images observed by COR2/STEREO-A (a, e, i) and synthesized from the electron number density which is treated to be the same as the proton number density in the time-evolving simulation results with the decomposed energy equation (b, f, j) ranging from 2.5 to 15 $R_s$, together with the simulated proton number density $N, (10^3~\mathrm{cm^{-3}})$ (c, g, k), and radial plasma velocity $V_r \, (\mathrm{km~s^{-1}})$ ranging from near solar surface to 15 $R_s$. The orange circle marks the position of 2.5~$R_s$. The first, second, and third rows correspond to the 170th, 340th, and 510th hours of the time-evolving coronal simulation during CR 2296.}\label{pBmeridianevolution}
\end{figure*}
Figure~\ref{pBmeridianevolution} shows the white-light pB images observed by  COR2/STEREO-A (a, e, i) and synthesized from the electron number density, which is treated
to be the same as the proton number density in the time-evolving coronal simulation (b, f, j), along with the simulated plasma density (c, g, k) multiplied by $\left( r \big/ R_s\right)^2$ with $r$ denoting the heliocentric distance and the simulated radial velocity (d, h, i). It reveals that the bright structures observed around $80^\circ~\rm N$ in the western hemisphere and around $80^\circ~\rm S$ in the eastern hemisphere are captured by the simulated pB images. The bright structures observed around $45^\circ~\rm S$ and $60^\circ~\rm S$ in the eastern and western hemispheres are captured in the simulation result at the 340th hour. At the 510th hour, the three bright structures observed in the western hemisphere are also largely captured in the simulated pB image, although the simulated bright pB structure around $30^\circ~\rm N$ and $60^\circ~\rm S$ are closer to the equator. The bright structure around $60^\circ~\rm S$ observed in the eastern hemisphere is also captured in the observation at the 510th hour. Regarding the discrepancy between observations and simulation results, for example, the simulated pB images covering the southern polar region are connected, whereas the bright pB structures in the corresponding high-latitude region are separated by the southern pole. The discrepancy may be attributed to the fact that the inner-boundary magnetic field in high latitudes, which are not accurately observed, is not well consistent with the real magnetic field configurations. Additionally, limitations of the synoptic magnetograms include the fact that the magnetic field at different longitudes is observed at different times, and that the empirically defined heating source terms cannot mimic the complex thermodynamic processes during solar maxima, which may also lead to discrepancies between simulated and observed pB results. Further considering the simulated plasma density and radial velocity distribution, we can conclude that the coronal model COCONUT, adopting a decomposed energy equation, successfully creates the low-density, high-speed (LDHS) and the high-density, low-speed (HDLS) structured coronal and solar wind, and distribution of the LDHS flows, usually manifested as bright structures in pB observations, are basically in agreement with observations. Further improvement may be achieved by adopting more realistic synchronized magnetograms and more physically consistent heating and acceleration source terms to constrain the coronal model.

\begin{figure*}[htpb]
\begin{center}
  \vspace*{0.01\textwidth}
    \includegraphics[width=0.8\linewidth,trim=1 1 1 1, clip]{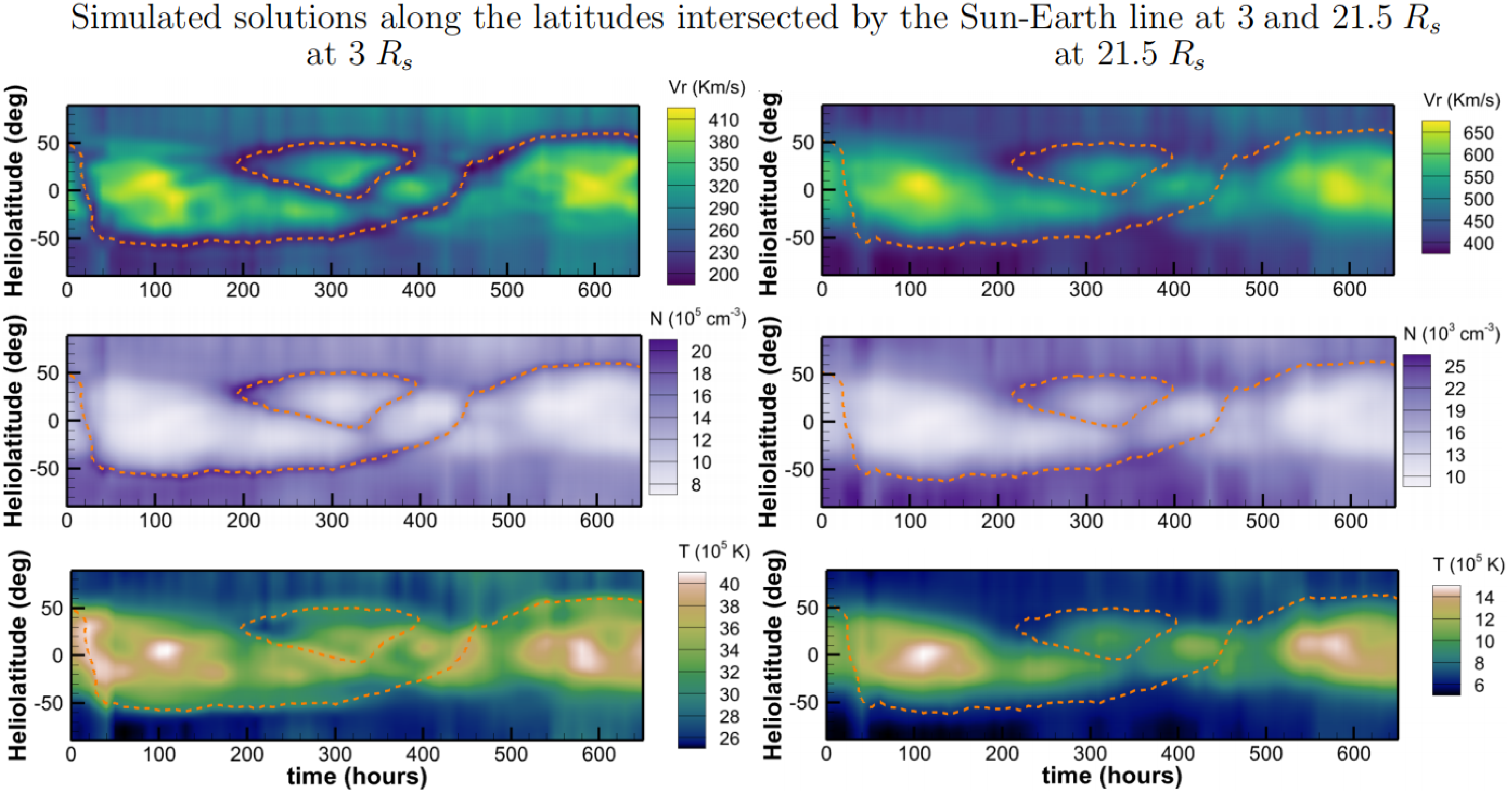}
\end{center}
\caption{Timing diagrams of simulated radial velocity $V_r$ ($\rm km~s^{-1}$, top), plasma number density ($\rm 10^5 ~ cm^{-3}$ at 3 $R_S$ and $\rm 10^3 ~ cm^{-3}$ at 0.1 AU, middle), and temperature ($\rm 10^{5}~K$, bottom) along the latitudes intersected by the Sun–Earth line at 3 $R_s$ (left) and 0.1 AU (right). The dashed orange lines indicate the MNLs derived from the time-evolving coronal simulations calculated by COCONUT using the decomposed energy equation.}\label{Timingevolution}
\end{figure*}
The diagrams in Figure~\ref{Timingevolution} are synthesized from a series of time-evolving simulation results, with a cadence of one result per 10 hours, along the longitude at $3~R_s$ and 0.1 AU passed by the Sun-Earth line. It reveals that the simulated HDLS flows are mainly concentrated around the MNLs. The distributions of the HDLS and LDHS flows at $3~R_s$ and 0.1~AU are essentially consistent. It is noticed that the magnitude of the simulated temperature is proportional to the radial velocity \citep{Elliott2012,Pinto_2017,Licaixia2018,wang2025sipifvmtimeevolvingcoronalmodel}, with the fast solar wind ($V_r>550~\rm km~s^{-1}$) at 0.1 AU occupied with temperatures ranging between 1 and 1.6 million Kelvin (MK), and with the slow solar wind ($V_r<450~\rm km~s^{-1}$) occupied with temperatures ranging between 0.5 and 0.7\;MK. It reveals that this model produces a reasonable latitudinal distribution of the LDHS fast flows and the HDLS slow flows at 0.1 AU.

\section{Concluding remarks}\label{sec:Conclusion}
Recently, the implicit MHD coronal model COCONUT has been developed into a highly-efficient time-evolving coronal model capable of performing faster-than-real-time coronal evolutions with only several dozen CPU cores \citep{Wang2025_FirsttimeevolvingCOCONUT}. In \cite{wang2025COCONUTMayEvent}, the algorithm, like PP measures implemented to the plasma density and thermal pressure, further improves the numerical stability of COCONUT, enabling the COCONUT to mimic the evolution of the coronal structures during solar maximum with a filter appropriately implemented to the magnetograms to constrain the magnetic field strength in the low corona not obviously exceeding 30 Gauss. In this paper, we further enhance the numerical stability of COCONUT by adopting a decomposed energy equation, in which the decomposed energy, comprising internal and kinetic energy, is updated during coronal simulations. In this method, the thermal pressure is derived by subtracting the kinetic energy from the decomposed energy. It avoids the subtraction of a large magnetic field energy from the total energy to obtain a significantly small thermal pressure, the magnitude of which can be smaller than the discretization error in the magnetic energy and usually leads to a nonphysical negative thermal pressure. 
Since the decomposed energy equation can no longer be expressed strictly in conservative form, even in ideal MHD, we additionally introduce an approximate dissipation term into the energy component of the HLL Riemann solver when computing the inviscid flux to further improve numerical stability. Consequently, COCONUT becomes more numerically stable and is able to calculate coronal evolutions during solar maxima even with the magnetic field strength near the active region in the low corona reaching up to 100 Gauss. This method is much easier to implement in an existing MHD coronal model, while also having the potential to be as numerically stable as the extended magnetic field decomposition method recently proposed in \cite{wang2025sipifvmtimeevolvingcoronalmodel}.

The nearly identical simulation results for a solar minimum case calculated by COCONUT with the decomposed and traditional full energy equations, with average relative differences in the plasma density, the velocity and the magnetic field strength being only $0.57\%$, $0.69\%$ and $0.21\%$, validate the reliability of this version of COCONUT adopting the decomposed energy equation. In the increasing phase case, the average relative differences in plasma density, velocity, and magnetic field strength are up to $5.61\%$, $6.14\%$, and $1.69\%$, respectively, demonstrating that the decomposed energy equation differs from the traditional method for coronal simulations with complex and strong magnetic field configurations. 
Furthermore, the nearly identical simulation results obtained in most regions of the 2D MHD vortex and rotor cases, except for moderate discrepancies near discontinuities, using both the traditional method and the algorithm proposed in this paper demonstrate that the decomposed energy equation, together with the HLL solver incorporating the additional dissipation term in the energy component, can still maintain consistency with the full energy equation results in most regions, even in the presence of numerous complex discontinuous structures. These 2D benchmark tests also indicate that the discrepancies, which are primarily concentrated around discontinuities, can be mitigated by using a finer mesh near the discontinuous structures. Additionally, successfully calculating the solar maximum simulation with magnetic field strength exceeding 50 Gauss, in which COCONUT, adopting the traditional full energy equation, crashed, further highlights the improvement in numerical stability made by the decomposed energy equation. Consistency with the coronal observations further demonstrates that COCONUT, by adopting the decomposed energy equation, is promising for future implementation in practical applications, such as daily space weather forecasting. 

Generally, from both physical and mathematical perspectives, the internal energy, the decomposed energy (the sum of internal and kinetic energies), and the total energy (the sum of internal, kinetic, and magnetic energies) are strictly equivalent. The internal energy equation can even be regarded as a fundamental form directly derived from the first law of thermodynamics at least within the truncation error. However, in the ideal MHD equations, evolving the internal energy typically requires explicitly computing the volume integration of $p \, \nabla \cdot \mathbf{v}$, while evolving the decomposed energy equation requires explicitly computing the volume integrations of $\mathbf{B} \cdot (\mathbf{v} \cdot \nabla \mathbf{B})$ and $\mathbf{v} \cdot (\mathbf{B} \cdot \nabla \mathbf{B})$. All of these computations can be avoided when evolving the total energy equation.  
Therefore, when numerical discretization errors are taken into account, discrepancies among the simulation results obtained from these three forms of energy equations arise in the presence of discontinuities and shocks, where accurately reconstructing the distribution of terms mentioned above within discretized cells is particularly challenging. 
Evolving the total energy ensures total energy conservation in the discrete sense. In contrast, evolving only the internal energy in numerical simulations often leads to discrepancies among the thermal pressure, momentum equation, and induction equation near discontinuities, thereby violating momentum and energy conservation.
In comparison, although decomposing the energy in numerical simulations does not strictly guarantee total energy conservation, it preserves consistency between the kinetic energy and the momentum equation, making it a reasonable compromise in terms of numerical consistency.
Additionally, in coronal simulations, addressing numerical instabilities in the low-$\beta$ regions is typically the more pressing issue. Therefore, in such cases, it is more appropriate to evolve the decomposed energy equations.

Although it has many merits, there are still many challenges to address to further improve the current COCONUT. Adaptive mesh refinement technique, as well as high-order accurate algorithms, are required to balance high accuracy and required efficiency for simulating transient events with high fidelity in practical applications, such as CME simulations triggered by sunspot rotations. synchronized magnetograms, in which the magnetic field at different longitudes is observed simultaneously, with magnetic field accuracy in polar regions significantly improved, are required to further enhance the consistency between simulations and observations. More realistic, physically based source terms for coronal heating and solar wind acceleration are necessary to enhance the reliability of coronal simulations involving highly complex magnetic and flow field configurations.

As seen from the validation of the present paper,
the improved COCONUT model is capable of modeling high solar activity phases, thereby enabling more reliable full solar cycle modeling. In the spirit of time-evolving modeling, we envision a number of possible future tasksa. First, we plan to utilize the vector magnetic field containing the tangential component to drive the temporal evolution of coronal structures and to apply
characteristic boundary conditions \citep{Hayashi_2021,Feng_2023,Tarr_2024,Kee_2025} to possibly improve physical consistency and numerical stability at the inner boundary. Then, we attempt to incorporate synchronized magnetograms, obtained through a series of time-dependent photospheric magnetograms,
including surface flux transport models \citep{Feng_2012,Hoeksema2020,Caplan_2025,Downs2025PSI} and artificial intelligence
based models \citep{Jeong_2020,Jeong_2025sft,Jeong_2025,Jeong_2025B}, to drive coronal-heliospheric evolutions with greater realism. In particular, a promising aspect is to generate the MHD process of a CME with a data-driven procedure in
the spherical wedge-shaped domain fitted to the  local solar eruptive region \citep{Jiang201605,Jiang2021,Jiang2023}, and subsequently combine the modeled MHD process of a CME in the wedge-shaped domain with the global MHD model \citep{Feng_2023}, so as to  self-consistently  develop a reliable data-driven CME modeling framework suitable for practical applications.

\textbf{\textit{Acknowledgments}}: 
This project has received funding from the European Research Council Executive Agency (ERCEA) under the ERC-AdG agreement No.~101141362 (Open SESAME). Neither the European Union nor the granting authority can be held responsible for them.
These results were also obtained in the framework of the projects FA9550-18-1-0093 (AFOSR), C16/24/010  (C1 project Internal Funds KU Leuven), G0B5823N and G002523N (WEAVE) (FWO-Vlaanderen), and 4000145223 (SIDC Data Exploitation (SIDEX), ESA Prodex).
This work is also supported by the National Natural Science Foundation of China (grant No.\ 42030204) and the BK21 FOUR program of the Graduate School, Kyung Hee University (GS-1-JO-NON-20242364).
E.H. is grateful to the Space Weather Awareness Training Network (SWATNet), funded by the European Union’s Horizon 2020 research and innovation program under the Marie
Skłodowska-Curie grant agreement No. 955620. The resources and services used in this work were provided by the VSC (Flemish Supercomputer Centre), funded by the Research Foundation – Flanders (FWO) and the Flemish Government. This work utilises data obtained by the Global Oscillation Network Group (GONG) program, managed by the National Solar Observatory and operated by AURA, Inc., under a cooperative agreement with the National Science Foundation. The data were acquired by instruments operated by the Big Bear Solar Observatory, High Altitude Observatory, Learmonth Solar Observatory, Udaipur Solar Observatory, Instituto de Astrof{\'i}sica de Canarias, and Cerro Tololo Inter-American Observatory. The authors also acknowledge the use of the STEREO/SECCHI data produced by a consortium of the NRL (US), LMSAL (US), NASA/GSFC (US), RAL (UK), UBHAM (UK), MPS (Germany), CSL (Belgium), IOTA (France), and IAS (France). This work also makes use of data obtained by LASCO C2/SOHO, EIT/SOHO and AIA/SDO instruments.

\begin{appendix}
\section{2D benchmark tests}\label{sec:benchmark}
We also conduct the two-dimensional (2D) Orszag–Tang vortex and MHD rotor test problems to further validate the decomposed energy strategy and the HLL Riemann solver with an additional dissipation term in the energy equation.
The 2D Orszag–Tang vortex system exhibits many key features of MHD turbulence, with shocks and other discontinuities developing in the flow as time evolves \citep{orszag_tang_1979}. The 2D MHD rotor test \citep{BALSARA1999270} consists of a dense rotating disk embedded in a static background fluid, with a velocity tapering layer between the disk’s edge and the ambient medium, and is usually used to evaluate the propagation of strong torsional Alfv{\'e}nic waves. These two cases have been used as classical benchmark tests in many research works \citep{Stone_2008,CIUCA2020100042,LIU2025113795,WangSIPtheoriticalCME}.

In this section, the 2D test cases are simulated by three different versions of COCONUT. Version 1: Cases calculated by full energy equation use the standard HLL Riemann solver, where the dissipation term is defined in Eq.~(\ref{InviscidFluxHLL}), and the results are indicated by a superscript ``$^{\rm FullE}$". Version 2: Cases simulated by the decomposed energy equation method employ the standard HLL Riemann solver default, and the results are denoted by a superscript ``$^{\rm DecE~without~extra~disspation}$". Version 3: Cases solved with decomposed energy equation and the HLL Riemann solver with an additional dissipation term in the energy equation, as described in Eq.~(\ref{InviscidFluxLLFandHLL}) with $\alpha = 0.125$, are indicated by a superscript ``$^{\rm DecE~with~extra~disspation}$". 

All simulations are conducted in a $[0,1] \times [0,1]$ domain discretized into triangular cells, with mesh resolutions of $200 \times 200$ and $1000 \times 1000$. Similar to \cite{CIUCA2020100042}, the initial condition for the Orszag–Tang vortex test is defined as in Eq.~(\ref{vortexInitial}).
\begin{equation}\label{vortexInitial}
\rho=\frac{25}{36~\pi}\,,~p=\frac{5}{12~\pi}\,,~
u=-\sin(2~\pi~y)\,,~v=\sin(2~\pi~x)\,,~w=0\,,~
B_x=\frac{\sin(2~\pi~y)}{\sqrt{4~\pi}}\,,~B_y=-\frac{\sin(4~\pi~x)}{\sqrt{4~\pi}}\,,~B_z=0\,.
\end{equation}
For the MHD rotor case, we set the initial condition as in Eq.~(\ref{RotorInitial}).
\begin{equation}\label{RotorInitial}
B_x=\frac{2.5}{\sqrt{4~\pi}}\,,~B_y=0\,,~B_z=0\,,~w=0\,,~(\rho, u, v) =
\begin{cases}
\left(10,~ -\dfrac{(y - 0.5)}{r_0},~ \dfrac{(x - 0.5)}{r_0}\right), & \text{if } r < r_0 \\[8pt]
\left(1 + 9\lambda,~ -\lambda\dfrac{(y - 0.5)}{r},~ \lambda\dfrac{(x - 0.5)}{r}\right), & \text{if } r_0 < r < r_1 \\[8pt]
\left(1,~ 0,~ 0\right), & \text{if } r > r_1
\end{cases}\,.
\end{equation} Here $r_0=0.1$, $r_1=0.115$, $r=\sqrt{\left(x-0.5\right)^2+\left(y-0.5\right)^2}$ and $\lambda=\frac{r_1-r}{r_1-r_0}$. For both cases, periodic boundary conditions are applied, and the adiabatic index is set to $\gamma = \frac{5}{3}$.

\begin{figure*}[htpb]
\begin{center}
  \vspace*{0.01\textwidth}
    \includegraphics[width=0.9\linewidth,trim=1 1 1 1, clip]{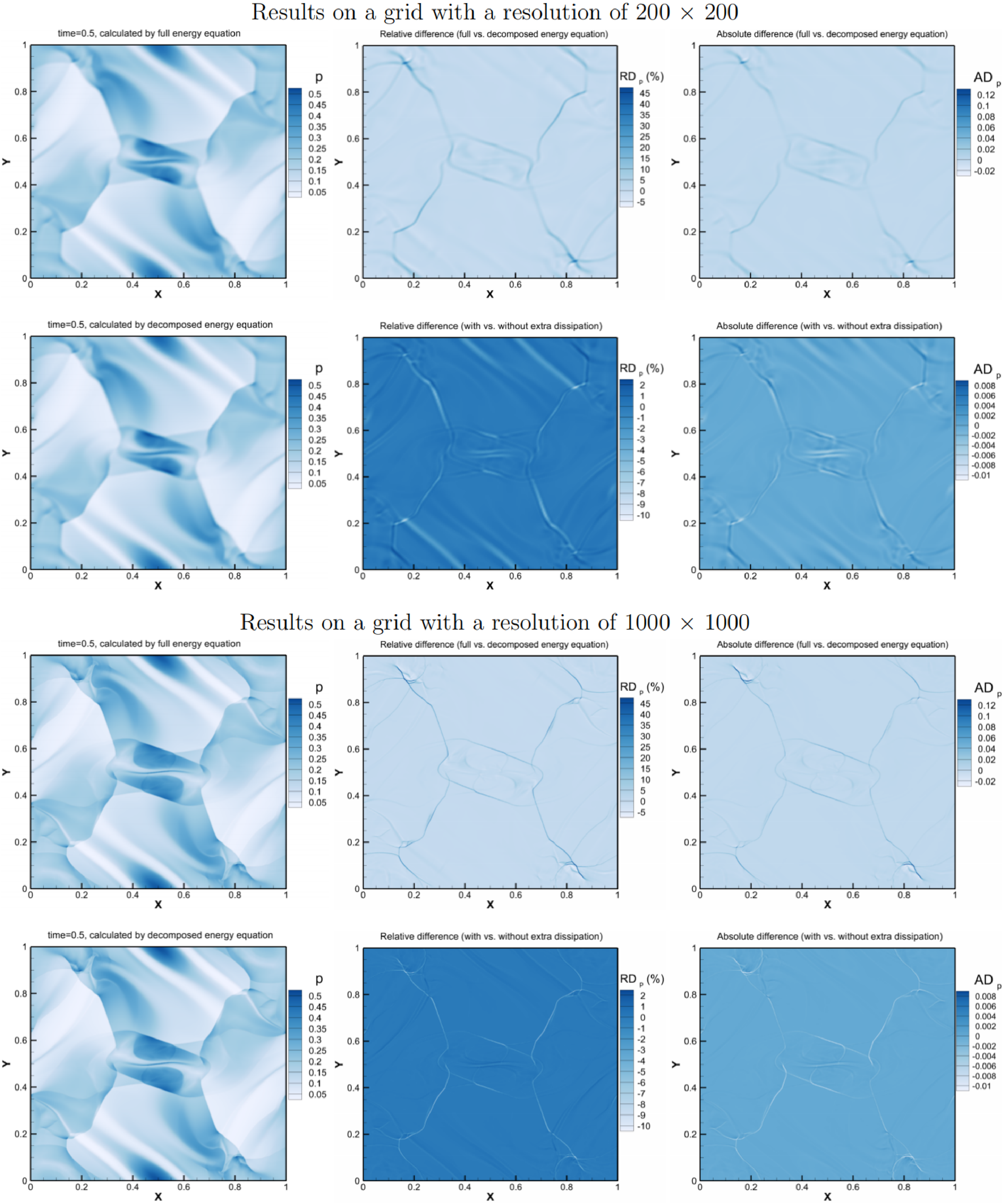}
\end{center}
\renewcommand{\thefigure}{A.~1}
\caption{Pressure distributions for the Orszag–Tang vortex problem at $t=0.5$, computed using the full energy equation, the decomposed energy equation, and the decomposed energy equation with an added dissipation term in the energy component of the HLL scheme, respectively. The first two and the last rows are calculated on the mesh with a resolution of $200\times200$ and $1000\times1000$, respectively. The middle and right columns display the relative and absolute differences: first and third rows compare the full and decomposed energy equations, while second and fourth rows compare the decomposed energy equation with and without the extra dissipation added to the energy component of the HLL scheme.}\label{p_OZ}
\end{figure*}

\begin{figure*}[htpb]
\begin{center}
  \vspace*{0.01\textwidth}
    \includegraphics[width=0.9\linewidth,trim=1 1 1 1, clip]{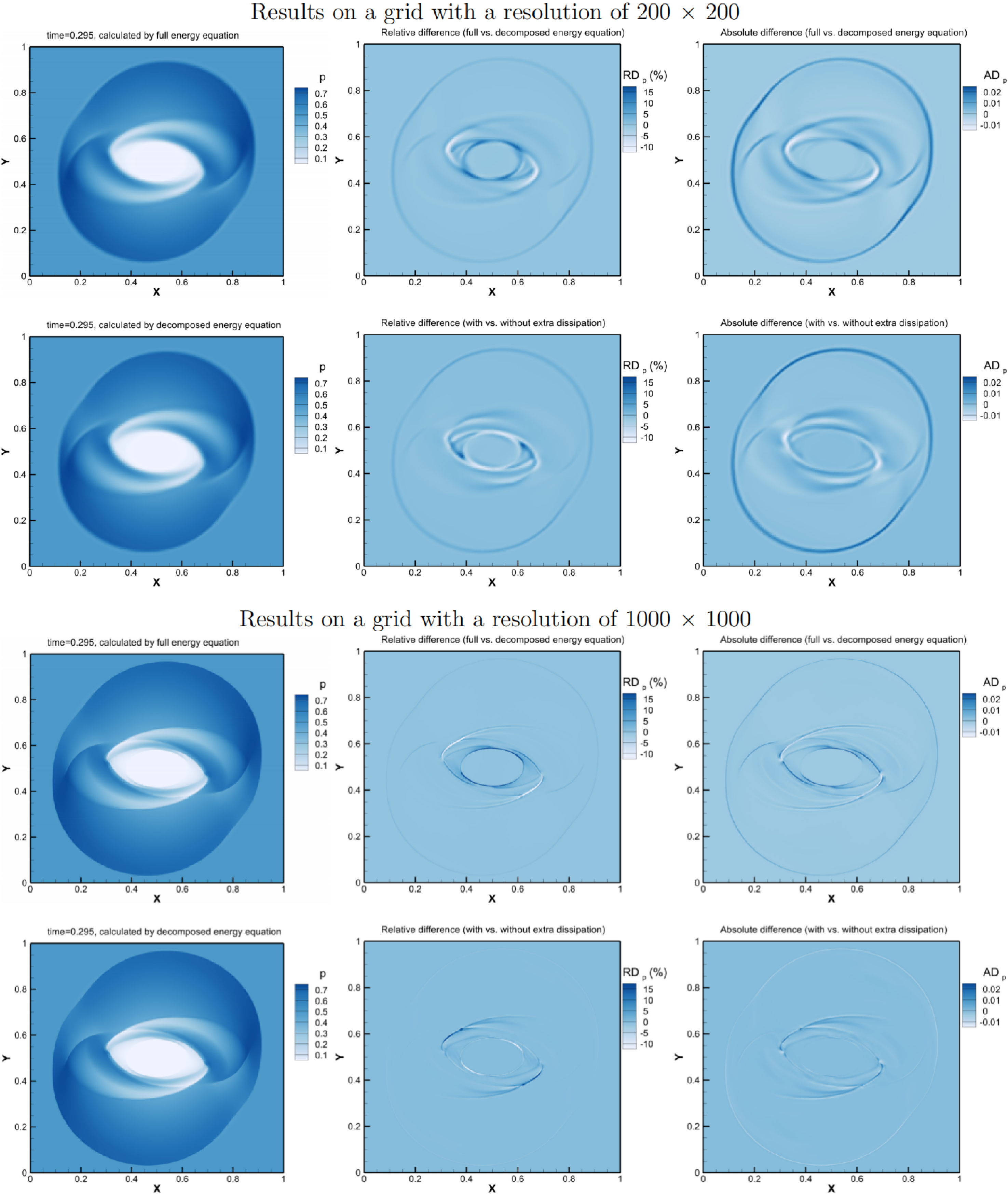}
\end{center}
\renewcommand{\thefigure}{A.~2}
\caption{Pressure distributions for the MHD rotor problem at $t=0.295$, computed using the full energy equation, the decomposed energy equation, and the decomposed energy equation with an added dissipation term in the energy component of the HLL scheme, respectively. The first two and the last rows are calculated on the mesh with a resolution of $200\times200$ and $1000\times1000$, respectively. The middle and right columns display the relative and absolute differences: first and third rows compare the full and decomposed energy equations, while second and fourth rows compare the decomposed energy equation with and without the extra dissipation added to the energy component of the HLL scheme.}\label{p_MHDRotor}
\end{figure*}

In Figures~\ref{p_OZ} and \ref{p_MHDRotor}, we present simulation results of the Orszag-Tang vortex and the MHD rotor problems at $t=0.5$ and $t=0.295$, respectively. These figures show pressure contours obtained from versions 1 and 2 of COCONUT, with the corresponding panels titled ``calculated by full energy equation” and “calculated by decomposed energy equation", respectively. These figures also show the relative differences (${\rm RD}_{p}$) and absolute differences (${\rm AD}_{p}$) in pressure calculated by the three versions of COCONUT.
For the panels labeled ``(full vs. decomposed energy equation)", the relative and absolute differences of pressure are defined as:
${\rm RD}_{p}=\frac{p^{\rm FullE}-p^{\rm DecE~without~extra~disspation}}{p^{\rm FullE}}$ and ${\rm AD}_{p}=p^{\rm FullE}-p^{\rm DecE~without~extra~disspation}$.
For the panels labeled ``(with vs. without extra dissipation)", the corresponding relative and absolute pressure differences are calculated as:
${\rm RD}_{p}=\frac{p^{\rm DecE~without~extra~disspation}-p^{\rm DecE~with~extra~disspation}}{p^{\rm DecE~without~extra~disspation}}$ and ${\rm AD}_{p}=p^{\rm DecE~without~extra~disspation}-p^{\rm DecE~with~extra~disspation}$.
It shows that the differences introduced by the decomposed energy formulation and the additional dissipation in the energy component of the HLL Riemann solver are negligibly small, with relative differences of less than $2\%$, except in the vicinity of very steep discontinuities. Moreover, discrepancies among the results calculated by different algorithms near discontinuities become less pronounced as the spatial resolution increases. 
The fact that noticeable differences are confined to very narrow regions around discontinuities further demonstrates that the decomposed energy equation and the added dissipation term do not affect the overall evolution of the MHD vortex and rotor structures.

\begin{figure*}[htpb]
\begin{center}
  \vspace*{0.01\textwidth}
    \includegraphics[width=0.9\linewidth,trim=1 1 1 1, clip]{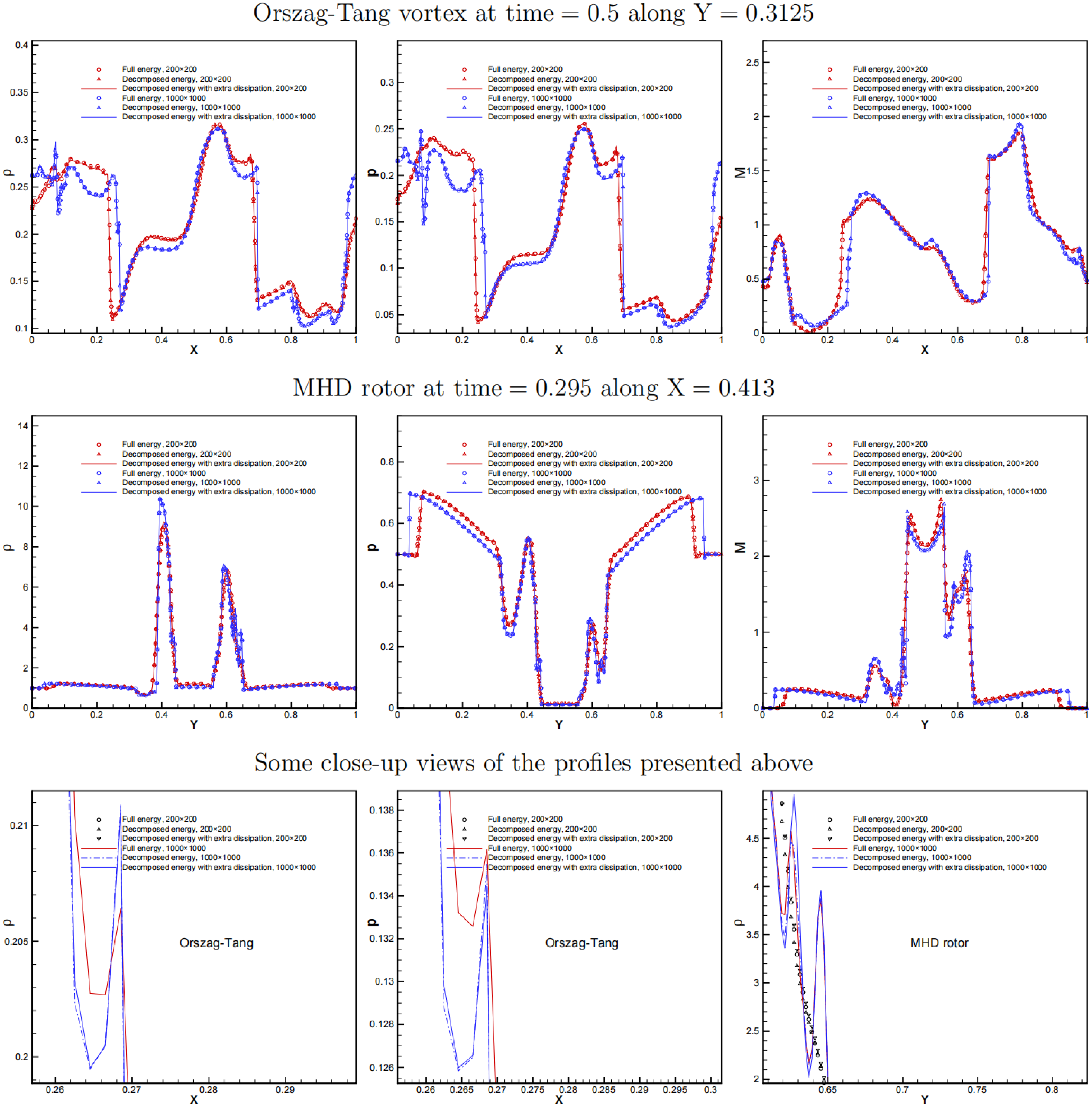}
\end{center}
\renewcommand{\thefigure}{A.~3}
\caption{Comparison of the full energy equation, the decomposed energy equation, and the decomposed energy equation with an added dissipation term in the energy component of the HLL scheme on grids with resolutions of $200\times200$ and $1000\times1000$ for the Orszag--Tang vortex at $t=0.5$ (first row) and the MHD rotor at $t=0.295$ (second row). The third row shows close-up views of the profiles in regions where relatively more noticeable differences are observed. This figure shows slices of dimensionless plasma density (left), pressure (middle), and Mach number (right, $M = \frac{|\mathbf{v}|\sqrt{\rho}}{\sqrt{\gamma p}}$) at $y = 0.3125$ for the Orszag--Tang vortex and at $x = 0.413$ for the MHD rotor, respectively.}\label{rho_p_Mach_OZandMHDRotor}
\end{figure*}

In Figure~\ref{rho_p_Mach_OZandMHDRotor}, we display the profiles of density $\rho$, thermal pressure $p$ and Mach number $\rm M=\frac{|\mathbf{v}|}{C_s}$ with $C_s$ being the sound speed along ${\rm Y}=0.3125$ at $t=0.5$ for the Orszag-Tang vortex cases and along ${\rm X}=0.413$ at $t=0.295$ for the MHD rotor tests, as did in \citep{CIUCA2020100042}. We also present close-up views of regions where the differences between results from different model versions are more pronounced. The results show that, at the same mesh resolution, the profiles of these variables from different model versions are in good agreement, indicating that the effect of the proposed algorithms is much smaller than that of the mesh resolution. Notably, two close-ups of the density and pressure profiles around ${\rm X}=0.265$ for the Orszag–Tang test, and one close-up of the density around ${\rm Y}=0.63$ for the MHD rotor test, reveal that the inflection structures calculated by the decomposed energy equation are more abrupt, indicating that the decomposed formulation can capture finer and more subtle features than the full energy equation.

\end{appendix}
\pagebreak

\bibliographystyle{aasjournal}
\bibliography{SIPandCOCONUT}

\end{document}